\documentclass{aa}   
\usepackage{graphicx}
\usepackage{txfonts}
\usepackage{lipsum}
\usepackage{subcaption}         
\usepackage{lscape}             
\usepackage{placeins}           
                                
\usepackage{xcolor}
\usepackage{hyperref}
\hypersetup{
    colorlinks=true,
    urlcolor=blue,
    linkcolor=red,
    breaklinks=true,
    citecolor=blue
}
\usepackage{etex}
\usepackage{scalerel}
\newcommand\HII{H\protect\scaleto{$II$}{1.2ex}}

\newcommand\Rpah{R$_{\mathrm{PAH}}$}

\newcommand\um{~$\mu$m}

\begin{document}
   \title{Uncovering the multi-scale structure of dust distribution in nearby galaxies} 



   \author{E.~Tanchon\inst{1}\fnmsep\inst{2} \and
 M.~Boquien\inst{1}\and
 J.~Chastenet \inst{3}\and
 D.~A.~Dale\inst{4} \and
 O.~V.~Egorov \inst{5}\and
 R.~Indebetouw \inst{6,7}\and
 R.~S.~Klessen\inst{8,9}\and
 S.~E.~Meidt\inst{3}\and
  D.~Pathak \inst{10} \and
 J.~Sutter \inst{11}\and
  D.~A.~Thilker \inst{12} \and
 A.~Amiri\inst{13} \and
 A.~T.~Barnes\inst{14}\and
 F.~Bigiel\inst{15} \and
 I.~S.~Gerasimov \inst{1}\and
 S.~C.~O.~Glover\inst{8}\and
 K.~Grasha\inst{16} \and
  K.~L.~Larson \inst{17} \and
 J.~C.~Lee \inst{18} \and
 H.-A.~Pan\inst{19}\and
 T.~G.~Williams\inst{20} \and
 the PHANGS collaboration
}
    \institute{Université Côte d'Azur, Observatoire de la Côte d'Azur, CNRS, Laboratoire Lagrange, France;
            \href{mailto:ella.tanchon@oca.eu}{ella.tanchon@oca.eu}
            \and 
            Ecole Normale Supérieure de Lyon, Lyon 69007, France
            \and 
            Sterrenkundig Observatorium, Universiteit Gent, Krijgslaan 281 S9, B-9000 Gent, Belgium
            \and
            Department of Physics and Astronomy, University of Wyoming, Laramie, WY 82071, USA
            \and 
            Astronomisches Rechen-Institut, Zentrum f\"{u}r Astronomie der Universit\"{a}t Heidelberg, M\"{o}nchhofstra\ss e 12-14, D-69120 Heidelberg, Germany
            \and
            Department of Astronomy, University of Virginia, 530 McCormick Road, Charlottesville, VA 22904, USA
            \and
            National Radio Astronomy Observatory, 520 Edgemont Road, Charlottesville, 22903, VA, USA
            \and
            Universit\"{a}t Heidelberg, Zentrum f\"ur Astronomie, Institut für Theoretische Astrophysik, Albert-Ueberle-Str. 2, 69120 Heidelberg, Germany
            \and
            Universit\"{a}t Heidelberg, Interdisziplin\"{a}res Zentrum f\"{u}r Wissenschaftliches Rechnen, Im Neuenheimer Feld 225, 69120 Heidelberg, Germany
            \and
            Department of Astronomy, Ohio State University, 180 W. 18th Ave, Columbus, Ohio 43210, USA
            \and 
            Whitman College Department of Astronomy, 345 Boyer Avenue, Walla Walla Wa, 99362
            \and
            Department of Physics and Astronomy, The Johns Hopkins University, 3400 North Charles Street, Baltimore, MD 21218, USA
            \and
            Department of Physics, University of Arkansas, 226 Physics Building, 825 West Dickson Street, Fayetteville, AR 72701, USA
            \and 
            European Southern Observatory (ESO), Karl-Schwarzschild-Stra{\ss}e 2, 85748 Garching, Germany
            \and
            Argelander-Institut für Astronomie, University of Bonn, Auf dem Hügel 71, 53121 Bonn, Germany
            \and
            Research School of Astronomy and Astrophysics, Australian National University, Canberra, ACT 2611, Australia
            \and
            AURA for the European Space Agency (ESA), Space Telescope Science Institute, 3700 San Martin Drive, Baltimore, MD 21218, USA
            \and
            Space Telescope Science Institute, 3700 San Martin Drive, Baltimore, MD 21218, USA
            \and
            Department of Physics, Tamkang University, No.151, Yingzhuan Road, Tamsui District, New Taipei City 251301, Taiwan
            \and
            UK ALMA Regional Centre Node, Jodrell Bank Centre for Astrophysics, Department of Physics and Astronomy, The University of Manchester, Oxford Road, Manchester M13 9PL, UK
            }

    \abstract
    {High-resolution JWST-MIRI images now allow us to resolve in great detail the multi-scale nature of the emission in nearby star-forming galaxies, from compact star-forming regions to large-scale diffuse emission, giving new insights into dust emission, its composition, and the surrounding interstellar medium (ISM).} 
    {We aim to understand at which scale the different processes driving dust emission in mid-infrared (7.7--21\um{}) wavelengths take place and if we can disentangle dense regions' emission from emission linked to a more diffuse component.}
    {We use and enhance the constrained diffusion decomposition (CDD) algorithm, an alternative to the wavelet transform decomposition, to disentangle the emission coming from compact regions from the emission originating from diffuse sources. This allows us to cleanly quantify the mid-IR spectral properties of the ISM at intervals within a continuum of physical scales.}
    {We find a transition scale of PAH emission around $300$~pc, with weaker PAH fraction at smaller scales, highlighting the destruction of PAHs in \HII{} regions. We also show variations in the PAH fraction in different morphological environments, with a smaller fraction in bright and star-forming environments. Studying and comparing the probability distribution functions (PDFs) of \HII{} regions and diffuse ISM  with the PDFs at different scales, we find a similar separation scale around $200$~pc at which we observe a transition from a power-law PDF for dense structures to a log-normal one for the diffuse ISM.}
    {We infer the destruction of PAHs in nebular environments and the existence of a transition scale between stellar feedback in nebular regions and heating in the diffuse ISM. We also show that the CDD enables future application to study physical scales of emission.}


   \maketitle
   \nolinenumbers

\section{Introduction}
A detailed understanding of star and dust formation, dust composition, and the interactions between dust and the interstellar medium (ISM), together with their characteristic timescales, is fundamental to building a complete view of galactic and stellar evolution. The ISM contains different structures on various scales \citep{Elmegreen_2001, thilker23, schinnerer_molecular_2024, bazzi25}, from the seeds of star formation: small dense clouds, \citep{bergin_cold_2007}, to large and diffuse structures of neutral gas \citep{Walter2008}, shaped by star formation, stellar feedback and environmental conditions \citep[e.g.,][]{sun18, Meidt_2021, Meidt_2023}. 

When massive and hot OB stars form in molecular clouds, they heat and ionise the surrounding gas, creating \HII{} regions. These can then be traced and observed in optical wavelengths using hydrogen recombination lines (e.g. H$\alpha \sim 6563$~\AA{} or H$\beta \sim 4861$~\AA{}; \citealt{Osterbrock_2006}). During the stars' lifetimes, \HII{} regions evolve and expand \citep{Klassen_2017} due to stellar feedback \citep[e.g. stellar winds, radiation pressure, protostellar jets, gas heating,][]{Agertz_2013, Barnes_2021,Barnes_2022,Pathak_2025}. The surrounding gas and dust reprocesses the UV light emitted by young, massive stars into optical, infrared, or mid-infrared (MIR) light \citep{Draine_2011}. Dust reprocesses on average $\sim30$\% of starlight into the infrared \citep[IR;][]{Bernstein2002} and its composition governs the ISM heating and attenuation. Dust is also a key component of ISM emission models as multiple spectral features are produced by different types of dust, with varying physical properties such as chemical composition, size, temperature and charge. 
Therefore, studying the properties of the IR emission is a way to recover information about the powering cluster and the dust composition in the ISM \citep{draine_infrared_2007, Calzetti_2009, Gregg_2025}.

Dust is composed of both large amorphous grains such as silicates or oxides (SiO$_2$, MgO, etc.) and smaller carbonaceous grains such as polycyclic aromatic hydrocarbons \citep[PAHs;][]{Tielens_2008, Draine_2011, Li_pah_2020}. Though the process of PAH formation is not completely determined \citep{Tokunaga2025}, recent studies show that PAHs are mostly formed in the atmosphere of evolved stars, rich in carbon \citep{Latter_1991}, but some PAH precursors have also been found in dense structures and molecular clouds \citep{Burkhardt_2021, Sandstrom_2010}. Another popular scenario is that PAHs are formed through the shattering of large dust grains by UV radiation or shocks \citep[e.g.][]{Wiebe_2014}. PAH destruction is believed to be governed by the radiation field and high intensity of ionising photons \citep{Montillaud_2013}, or by fragmentation due to electronic and atomic interactions \citep{Micelotta_2010,Micelotta_2011,Bocchio_2012}.

PAHs are known to have multiple emission features spanning the near to mid-IR, with some of the most prominent features at: $3.3$~$\mu$m, $7.7$~$\mu$m, and $11.2$~$\mu$m \citep{Allamandola_1989}, making those wavelengths very interesting to trace the PAH mass fraction in interstellar dust and the intensity of the interstellar radiation field (ISRF). The PAH mass fraction has been shown to vary between 0\% to 5\% in different environments in local galaxies and in Magellanic clouds \citep{draine_infrared_2007, Aniano_2020,Sandstrom_2010, Paradis_2011,Chastenet_2023a, sutter_fraction_2023,Egorov_2025}. Since the IR emission from PAHs is sensitive both to the amount of dust and to the strength of the incident radiation field, its distribution reflects both the structure of the ISM and the location of strong radiation sources such as OB stars \citep{pathak_two-component_2023}. Studying the distribution of emission, we can understand and trace the environmental properties of galaxies and the different phases of the ISM.

Measuring the probability distribution function (PDF) of gas column density, which can be derived from dust extinction, allows one to study the different phases of the ISM. Studies focused on the Milky Way and Local Group galaxies \citep{Lombardi_2009, Schneider_2015} combined with high-resolution numerical simulations \citep{Koyama_2009, Fensch_2023} have shown the column density PDF is shaped to first order by both turbulence and gravity. At high column densities, e.g.\ in dense clouds or filaments and in star-forming regions, which are dominated by gravity, the PDF follows a power-law \citep{Hennebelle_2012}. For smaller column densities, tracing more diffuse gas, the PDF follows a log-normal distribution, which can be attributed to supersonic isothermal turbulence, i.e. random compression of fluid particles by turbulent velocity fluctuations \citep{Kritsuk_2007}. Furthermore, it has been demonstrated \citep{Federrath_2008} that the turbulent component of the PDF is really log-normal only when the turbulence contains incompressible modes. 

By studying MIR emission, we can infer both dust composition and emission properties in the ISM. Previous work on the Milky Way and nearby starburst galaxies has demonstrated that the PAH mass fraction in the ISM can be constrained using Spitzer IRAC 8\um{} and MIPS 24\um{} data, and that it is closely linked to metallicity \citep[see e.g.][]{Engelbracht_metallicity_2005, Engelbracht_metallicity_2008}. Studies of column densities PDFs \citep{Berkhuijsen_2008, Corbelli_2018, pathak_two-component_2023} show how PDF can be a tool to identify gravity or turbulence dominated regions and therefore dense or diffuse regions in the ISM. However, both PAH and PDF studies have been limited by the resolution and sensitivity of telescopes.

JWST now provides a spatial resolution an order of magnitude sharper than Spitzer and allows one to trace PAHs using either spectroscopy \citep[e.g.][]{Pdr4all_1, Pdr4all_2, Pdr4all_3} or photometry \citep{JWST_first_1,JWST_first_2, Chastenet_2023a, Gregg_2024}. However, using spectroscopy, with either the MIRI IFU or with NIRSpec, is observationally costly if one wants to map large areas, owing to the small field-of-views of these instruments. Therefore, large-scale studies of PAHs in nearby galaxies have primarily focused on photometry using MIRI and NIRCam. These contain photometric filters centred around several of the strong PAHs emission lines at $z=0$, enabling the study of PAH emission in nearby ($\sim 10$~Mpc) galaxies. Furthermore, we are able to resolve star-forming regions in nearby galaxies and to study the PDF of IR emission in both dense and diffuse regions.

Following previous work by \cite{Chastenet_2023a}, \cite{Egorov_2023}, \cite{sutter_fraction_2023} and \cite{pathak_two-component_2023} we use the MIR observation of 19 galaxies with JWST from the PHANGS-JWST Cycle 1 Treasury \citep{Lee_2023} to understand how the infrared emission is related to the different structures and environments in galaxies. Using a constrained diffusion algorithm, we separate different emission scales and analyse how the PAH and warm dust emission varies with the scale.

This paper is organized as follows: In Section \ref{data}, we present the data and the decomposition algorithm used to conduct our study. In Section \ref{sec:rpah_sec}, we present PAH maps at different scales and in different environments, while Section \ref{PDF} shows the PDF at each scale and how it can be a tool to distinguish emission from inside or outside \HII{} regions. Section \ref{discussion} presents our results in regards to previous studies. Finally, Section \ref{conclusion} summarizes our results and analysis.  

\section{Data and Methods}
\label{data}
For this study, we use observations of 19 nearby star-forming galaxies observed with JWST during the PHANGS-JWST Cycle 1 Treasury program \citep[Program 2107, PI Lee;][]{Lee_2023} and with MUSE on the VLT as part of the PHANGS-MUSE survey \citep{emsellem_phangs-muse_2022}. The properties of these galaxies are detailed in Table~\ref{tab:MUSE_carac}.
\begin{table*}[!htbp]
    \begin{center}
    \begin{tabular}{l|cccccc}
        \hline
        \hline
        Name & Distance$^{(a)}$ & $\log(M_{\star})^{(b)}$ & $\log({\rm SFR})^{(c)}$ &  PA$^{(d)}$ & i$^{(d)}$ & PSF $^{(e)}$ \\
        & [Mpc] & [$M_{\odot}$] & [$M_{\odot}yr^{-1}$] & [deg] & [deg] & [pc] \\
        \hline
        IC5332 & \phantom{0}9.0 & \phantom{0}9.67 & $-1.30$ &  \phantom{0}74.4 & 26.9 &37.13  \\
        NGC0628 & \phantom{0}9.8 & 10.34 &$-0.35$ &  \phantom{0}20.7 & \phantom{0}8.9 &40.55  \\
        NGC1087 & 15.9 & \phantom{0}9.93 & \phantom{$-$}0.26 & 359.1 & 42.9 &65.32 \\
        NGC1300 & 19.0 & 10.62 & $-0.08$ & 278.0 & 31.8 &78.26 \\
        NGC1365* & 19.6 & 10.99 & \phantom{$-$}0.35 &  201.1 & 55.4 &80.64 \\
        NGC1385 & 17.2 & \phantom{0}9.98 & \phantom{$-$}0.44 & 181.3 & 44.0 &70.96 \\
        NGC1433 & 18.6 & 10.87 & $-0.25$ &  199.7 & 28.6 &76.77 \\
        NGC1512 & 18.8 & 10.71 & $-0.19$ &  261.9 & 42.5 &77.60\\
        NGC1566* & 17.7 & 10.78 & \phantom{$-$}0.50 &  214.7 & 29.5 &72.89 \\
        NGC1672* & 19.4 & 10.73 & \phantom{$-$}0.79 &  134.3 & 42.6 &79.95 \\
        NGC2835 & 12.2 & 10.00 & $-0.27$ &  \phantom{00}1.0 & 41.3 &50.35 \\
        NGC3351 & 10.0 & 10.36 & $-0.24$ &  193.2 & 45.1 &41.04\\
        NGC3627* & 11.3 & 10.83 & \phantom{$-$}0.71 &  173.1 & 57.3 &46.65\\ 
        NGC4254 & 13.1 & 10.42 & \phantom{$-$}0.52 &  \phantom{0}68.1 & 34.4 &53.98 \\ 
        NGC4303* & 17.0 & 10.52 & \phantom{$-$}0.69 &  312.4 & 23.5 &70.01 \\
        NGC4321 & 15.2 & 10.75 & \phantom{$-$}0.48 &  156.2 & 38.5 &62.68 \\
        NGC4535 & 15.8 & 10.53 & $-0.01$ &  179.7 & 44.7  &64.99 \\
        NGC5068 & \phantom{0}5.2 & \phantom{0}9.40 & $-0.74$ &  342.4 & 35.7  &21.43 \\
        NGC7496* & 18.7 & 10.00 & $-0.29$ &  193.7 & 35.9 &77.14 \\ 
    \end{tabular}
    \caption{Parameters of the galaxies in our study. Galaxies with a $^*$ have AGN (Active Galaxy Nuclei) in their center. (a)~From the compilation of \cite{anand_distances_2021}. (b)~Derived by \cite{leroy_phangs-alma_2021}, using GALEX UV and WISE IR photometry, following a similar methodology to \cite{Leroy_2019}. (c)~Derived using the extinction-corrected H${\alpha}$ measurements from \cite{belfiore_calibration_2023}. (d)~From \cite{Lang_2020}, based on CO(2–1) kinematics. (e)~Size of the Point Spread Function (PSF) at 0.85" in parsec.}
    \label{tab:MUSE_carac}
    \end{center}
\end{table*}
To be observable by ALMA and MUSE, all targets have a declination $-75^\circ \leq \delta \leq +25^\circ$, are nearby (5~Mpc $\leq$ D $\leq$ 20~Mpc), have low to moderate inclination ($i < 75^\circ$) to limit the effects of extinction and line-of-sight confusion and facilitate the identification of individual star-forming sites, and are massive star-forming galaxies with stellar mass $\log(M_{\star}/M_{\odot}) \gtrsim 9.75$ and specific star formation rate $\log(\rm sSFR\,[yr^{-1}]) \gtrsim -11$. This sample includes spiral galaxies along the star-forming main sequence with various morphologies and some galaxies containing AGNs.

\subsection{JWST MIRI and NIRCam}
The PHANGS-JWST survey was designed to map and resolve the infrared emission from stellar clusters, young stellar populations, \HII{} regions, and molecular clouds. The survey uses imaging in eight bands, four from near-IR (NIRCam) and four mid-IR (MIRI) filters, from 2 to 21~$\mu$m. The coverage was optimised to cover as much as possible of the already observed data in optical spectroscopy with MUSE, requiring, on average, two pointings with NIRCam and four pointings with MIRI. We use MIRI data from four mid-IR filters: F770W, F1000W, F1130W, and F2100W, which are centred on 7.7~$\mu$m, 10~$\mu$m, 11.3~$\mu$m, and 21~$\mu$m, respectively. The JWST data were reduced using the \texttt{pjpipe} package \footnote{\href{https://pjpipe.readthedocs.io/en/latest/}{https://pjpipe.readthedocs.io/}} as described in \cite{williams_2024}.

The filters at 7.7~$\mu$m and 11.3~$\mu$m mostly trace PAH emission \citep{Allamandola_1989} while 10~$\mu$m and 21~$\mu$m trace the silicate dust and hot dust continuum. In low-density regions, the 21~$\mu$m filters also trace the stochastic heating of the grains \citep{Draine_2011}. We use the NIRCam F200W filter to trace and remove the underlying stellar emission in dust dominated bands (see Section \ref{sec:rpah_sec} of this paper and \citealt{sutter_fraction_2023}). We use convolved maps at 0.85\arcsec following the methodology of \citealt{aniano_common-resolution_2011}. This is larger than the F2100W filter PSF to account for the difference in Point Spread Function (PSF) in the different MIRI filters and to make sure all maps have the same resolution and that structures can be compared \citep{williams_2024}.
\subsection{PHANGS data products}
To conduct our analysis, we rely on other PHANGS data products, such as masks allowing us to separate different environments and regions in the galaxies.

\subsubsection{Nebular masks}
The nebular masks of \HII{} regions are defined by \cite{groves_phangs-muse_2023} using MUSE-H${\alpha}$ maps \citep[see][for a detailed presentation of the PHANGS-MUSE survey and its data products]{emsellem_phangs-muse_2022}. In this work, we only consider the subset of nebular regions that are flagged as \HII{} regions. Those regions are selected using the diagnostic curves described by \cite{Baldwin_1981} and applying a Signal-to-Noise ratio cut of 5. 

It is important to note that the definition of those regions is limited by the resolution of MUSE maps (which only resolves the largest regions) and that the \HII{} regions are generally smaller than the nebular regions defined in the catalogue (MUSE resolution is $\sim50$-$80$~pc, whereas \HII{} regions have typical sizes of $\sim10$~pc). The boundary of the regions is defined using a terminal gradient value of the H${\alpha}$ surface brightness to separate nebular regions from diffuse gas leading to smaller contours for \HII{} regions \citep{thilker_hiiplot_2000, Barnes_2023, Chandar_2025, Barnes_2025}. However, we assume in this work that those masks isolate \HII{} regions at high confidence, and that the surrounding regions correspond to mostly diffuse and neutral ISM. In our work, we refer to \HII{} regions (and masks) to describe all the nebulae from \cite{groves_phangs-muse_2023} catalogue (and associated spatial masks defining their borders) that were classified there as \HII{} regions in that work. We also refer to non-\HII{} regions to describe all the regions outside of the \HII{} masks.

\subsubsection{Environmental characterisation} \label{morph_mask}
\begin{figure}[!htbp]
   \centering
   \includegraphics[width=\columnwidth]{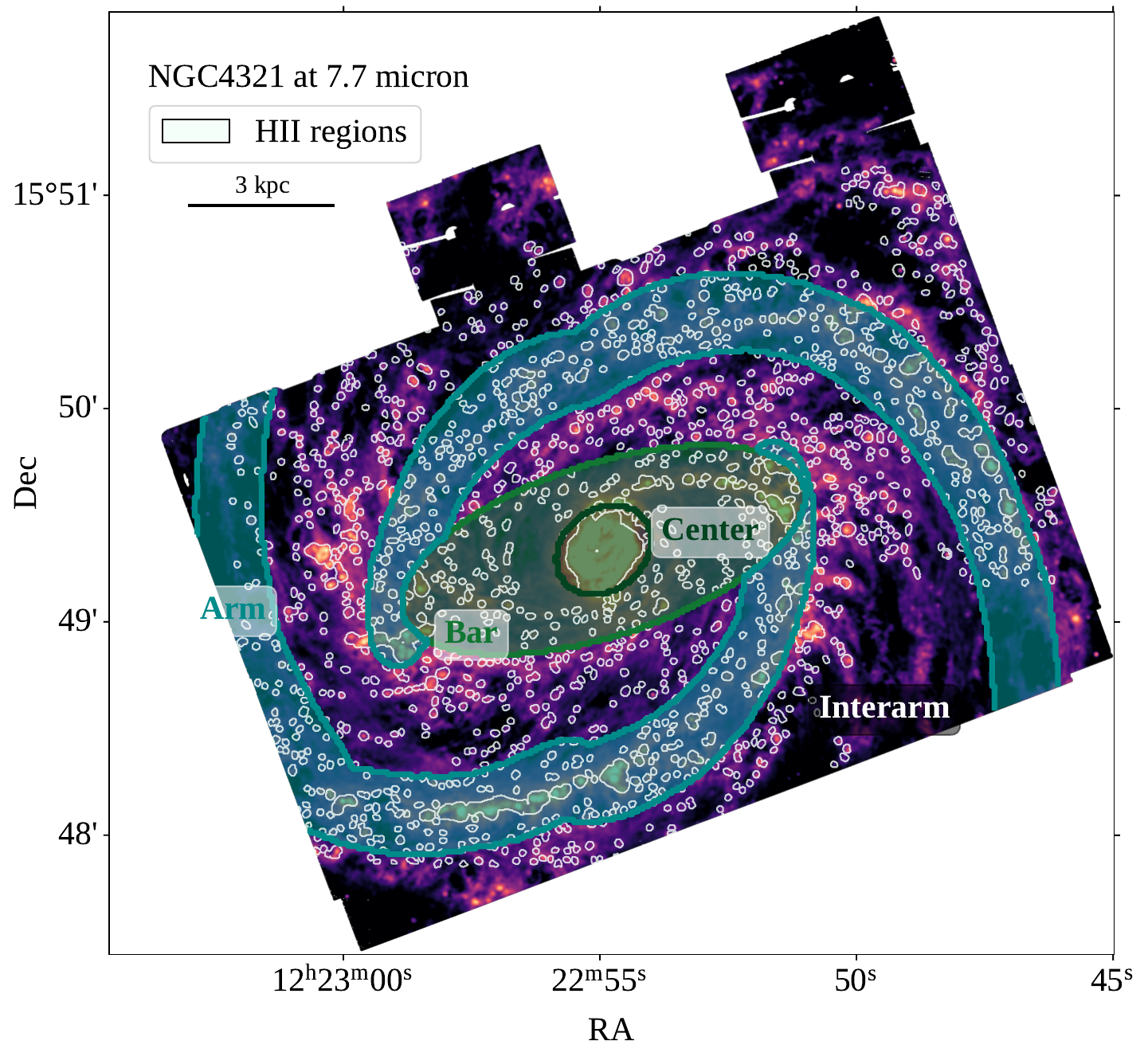}
      \caption{Environmental masks and \HII{} region masks for NGC4321 at 7.7\um{}. The \HII{} regions are highlighted with white contour and the environmental regions are shown by shaded areas.}
         \label{fig:masks}
\end{figure}

To study the dependence on environment of the emission, we use the environmental masks defined by \cite{querejeta_stellar_2021}. These masks were derived using Spitzer 3.6~$\mu$m data, and each pixel is associated with a dominant structure. The masks contain 4 main structures: centre, bar, spiral arms, and inter-arm regions (Fig.~\ref{fig:masks}), for each galaxy.
The inter-arm regions are complementary to the spiral regions. As some galaxies do not have spiral arms, the regions not corresponding to any other structures are labelled as disk. 

\subsection{Constrained Diffusion Decomposition Method}
To separate the different structures of the emission, we use the constrained diffusion algorithm, described in detail by \cite{li_multiscale_2022}. It decomposes an image into a set of maps containing the emission at different scales. To produce this set of images, this algorithm uses a modified version of the diffusion equation (see Appendix~\ref{app:dec_full} for a more detailed explanation of the method).

This method offers an alternative to the commonly used wavelet transform \citep{Starck_2003}, where the data are convolved to a set of Gaussian kernels of increasing size. The wavelet method produces artefacts around sharp edges and some areas of unnecessary negative values, making it less adapted in cases where local flux conservation at all scales is critical. On the contrary, with the constrained diffusion approach, the flux is conserved through the transformation, and the study of individual maps is meaningful when compared to the global flux of the full image (see \citealt{li_multiscale_2022} for an in-depth comparison).

\begin{table}[!htbp]
    \begin{center}
    \begin{tabular}{cc|cc}
        \hline
        \hline
        Scale & Size (pc) & Scale & Size (pc)\\
         number & &  number & \\
        \hline
        \phantom{0}1  & 100 -- 120 & 11 &  \phantom{0}619 -- \phantom{0}743 \\
        \phantom{0}2  & 120 -- 144 & 12 &  \phantom{0}743 -- \phantom{0}892 \\
        \phantom{0}3  & 144 -- 173 & 13 &  \phantom{0}892 -- 1070 \\
        \phantom{0}4  & 173 -- 207 & 14 & 1070 -- 1284 \\
        \phantom{0}5  & 207 -- 248 & 15 & 1284 -- 1541 \\
        \phantom{0}6  & 248 -- 298 & 16 & 1541 -- 1849 \\
        \phantom{0}7  & 298 -- 358 & 17 & 1849 -- 2219 \\
        \phantom{0}8  & 358 -- 430 & 18 & 2219 -- 2663 \\
        \phantom{0}9  & 430 -- 516 & 19 & 2663 -- 3195 \\ 
        10 & 516 -- 619 & 20 & 3195 < \phantom{0000} \\
    \end{tabular}
    \caption{Size of the logarithmically spaced bins used for the decomposition}
    \label{tab:scale_table}
    \end{center}
\end{table}   

We adapt the method to make it more specific for our study. The main adaptation we perform is on scale choices. The original algorithm uses logarithmic spacing with bins of size $2^n$ pixels, where $n$ is the scale number. As we want to study physical changes depending on environment and to compare their evolution in several galaxies, we define a new set of scales, different from the original one, in parsecs (see Table~\ref{tab:scale_table}), rather than pixels. This gives a more physical definition for each scale and it allows to easily compare the decomposition between galaxies.

\begin{figure*}[!htbp]
   \centering
   \includegraphics[width=\textwidth]{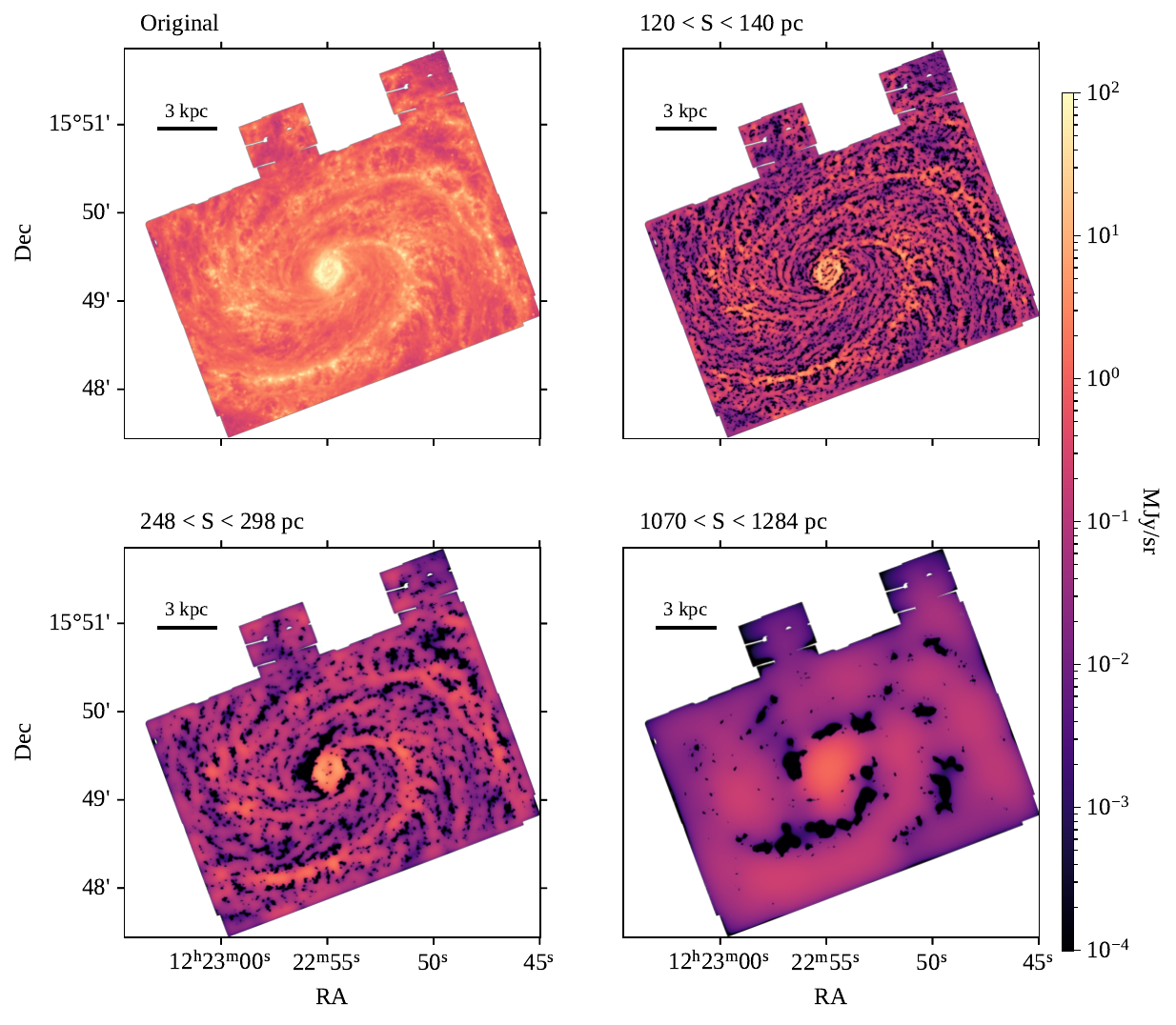}
      \caption{Example of the decomposition at different scales for NGC4321 at 7.7\um{}. An alternative version of this plot with all scales is shown in Appendix~\ref{app:dec_full}.}
         \label{fig:dec_scales}
   \end{figure*}

We produce a FITS cube containing the emission at each spatial scale with the same pixel scale as the original image. To analyse and characterize the emission, we performed the decomposition on a log-scale range. We chose our minimum scale to be above the PSF limit at our most distant target (NGC1365 at 19.6~Mpc, see Table~\ref{tab:MUSE_carac}) and set this minimal scale to 100~pc. Doing this, we lose information on structures of sizes lower than 100~pc. We also make sure that the flux is conserved by the decomposition and that we have the original intensity: $I(x,y)$ equal to $ \sum_n M_n(x,y)$, the sum of the intensity of each maps, as described previously (see Appendix~\ref{app:pow_spec}). We performed the decomposition on 20 log-spaced intervals. This choice of scales allow to have good sampling at small scales and reach high scales (see Figure~\ref{fig:dec_scales}) while keeping a reasonable computation time. We then apply the different masks (\HII{} regions and environmental) if needed.
\label{method}
   

\section{\texorpdfstring{\Rpah{}}{} Analysis}\label{sec:rpah_sec} 
To study the trend in PAH emission and of the PAH fraction, we need to define a strong and reliable indicator which account for both ionised and neutral PAH emission. Early studies with JWST by \cite{Chastenet_2023a} and \cite{Egorov_2023} considered:
\begin{equation}
    R_{\mathrm{PAH}} = \frac{\mathrm{F770W}_{\rm ss}+\mathrm{F1130W}}{\mathrm{F2100W}},
    \label{eq_rpah_ind}
\end{equation}
as a photometric tracer of $q_{\mathrm{PAH}}$, the PAH mass fraction, where F1130W and F2100W are the surface brightness in MJy~sr$^{-1}$ at 11.3 and 21\um{} and F770W$_{\rm ss}$ is the surface brightness in the band at 7.7\um{} with starlight subtraction (see Section~\ref{starsub} below). The reliability of \Rpah{} as a tracer of $q_{\mathrm{PAH}}$ was later confirmed by \cite{sutter_fraction_2023} (and Pathak et al. \textit{in prep}, Koziol et al. \textit{in prep}) by direct measurements of $q_{\mathrm{PAH}}$. This photometric tracer is consistent with the ratio of flux of 8\um{} over 24\um{} observations widely used in earlier work as a tracer of $q_{\mathrm{PAH}}$ with Spitzer data \citep[e.g.,][]{Engelbracht_metallicity_2005, Engelbracht_metallicity_2008, Sandstrom_2010}.

Deriving $q_{\mathrm{PAH}}$ maps necessitates full SED modelling, which, for our sample (see Section 3.2.1 of \citealt{sutter_fraction_2023}), would lead to a significant degradation of spatial resolution (18” for SPIRE 250 \um{} data). Furthermore, deriving $q_{\mathrm{PAH}}$ maps adds a dependence on the parameters and methods used to model dust emission (such as the size of dust grains, the ISRF value, etc.), adding bias to our data. Therefore, in the rest of our analysis, we only study \Rpah{}.
It is possible to use \Rpah{} as a proxy for the PAH fraction, assuming that: 
\begin{description}
    \item[\textbf{(I)}] At typical ISRF intensities and at large scales, dust and PAH emission dominate in the F770W and F1130W bands \citep{whitcomb_molecular_2023}. The cases where there is a high starlight contamination are taken into account by the starlight subtraction (see Section \ref{starsub}).
    \item[\textbf{(II)}] The sum of the F770W and F1130W bands is a good proxy for PAH emission as they contain emission features from ionized (F770W) and neutral (F1130W) PAHs \citep{draine_infrared_2007}.
    \item[\textbf{(III)}] RPAH does not vary as a strong function of radiation field in the diffuse ISM of normal star-forming disks \citep{Dale2001, Chastenet_2023a, sutter_fraction_2023}. Typical models \citep[e.g.][]{draine_infrared_2007} do not show substantial variations in the MIR bands and in the PAH emission when the ISRF intensity varies in the typical value range. Therefore, the MIR spectrum is not primarily affected by any effect other than PAH emission, and these changes do not impact the correlation between \Rpah{} and $q_{\mathrm{PAH}}$. 
\end{description}

\subsection{Starlight subtraction and \texorpdfstring{\Rpah{}}{} map construction}
\label{starsub}
In some galaxies, where the stellar surface density is high relative to the dust density, the F770W filter may contain a non-negligible contribution from starlight. To remove this contribution, we use the NIRCam F200W band (centred around 2\um{}) re-scaled and matched to the same same PSF to predict the starlight contribution in the F770W band. We use the F200W band over the F300M band for star subtraction as it has a higher signal to noise ratio and is less likely to be contaminated by hot dust. In this work, we use the scaling factor determined by \cite{sutter_fraction_2023} based on stellar emission models (CIGALE SED modelling \citealt{Boquien_2019}) that assume an \cite{Chabrier_2003} initial mass function, and are calibrated to take dust attenuation into account. We define for every galaxy:
\begin{equation}
    \mathrm{F770W}_{\rm ss} = \mathrm{F770W}-0.13 \times \mathrm{F200W}
\end{equation}
as the starlight-subtracted map in the 7.7\um{} band. We then perform the decomposition on the starlight-subtracted maps in order to build \Rpah{} maps for each scale. To study the value of the \Rpah{} distribution at each scale, we build ``decomposed'' maps of \Rpah{} with:
\begin{equation}
    R_{\mathrm{PAH}}^i = \frac{\mathrm{F770W}_{\rm ss}^i+\mathrm{F1130W}^i}{\mathrm{F2100W}^i},
    \label{eq_rpah}
\end{equation}
where FILTER$^i$ corresponds to the $i$-th scale of the image presented in Table~\ref{tab:scale_table}. We then apply the \HII{} and/or environmental masks on the decomposed maps if needed. 

\subsection{Median \texorpdfstring{\Rpah{}}{} value}\label{sec:rpah_med}
\begin{figure*}[!htbp]
    \centering
    \includegraphics[width=\textwidth]{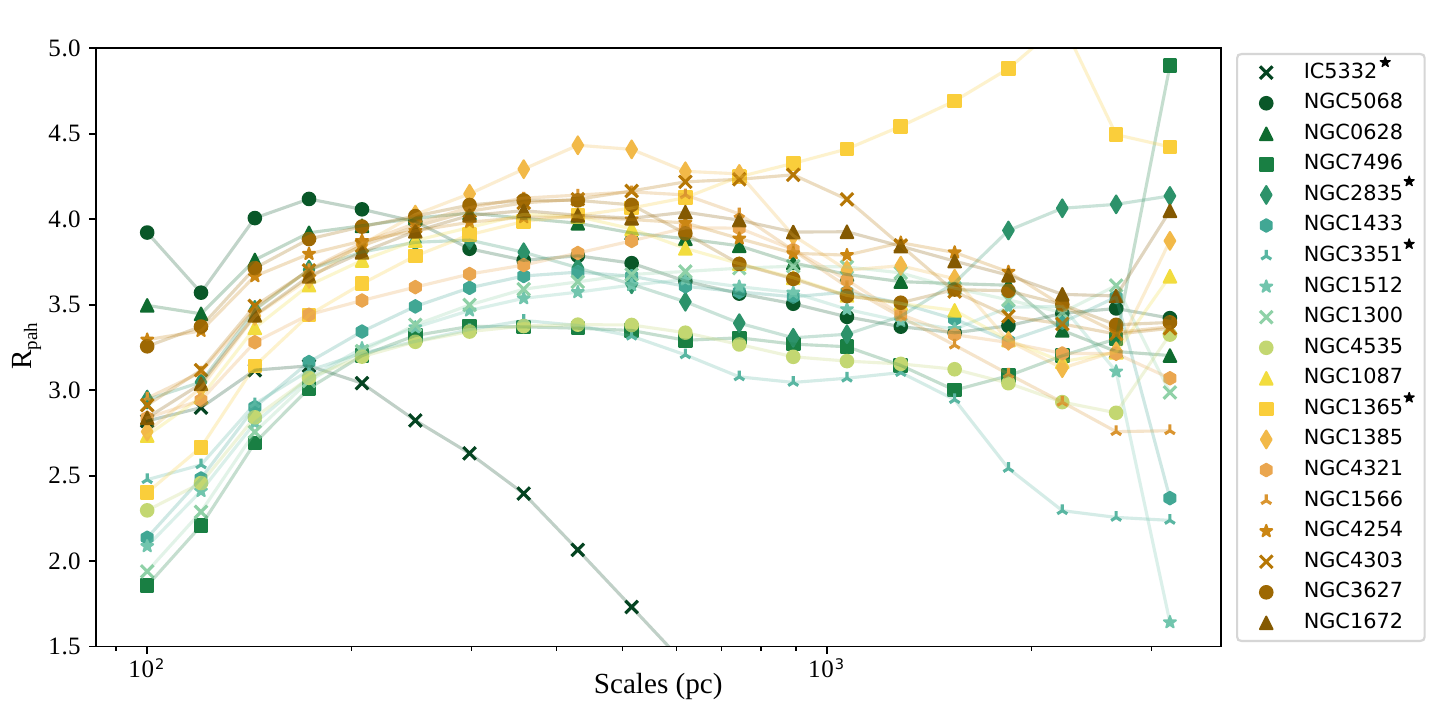}
    \caption{Median \Rpah{} values for all galaxies at each scale. Galaxies are colour-coded by SFR in increasing order with the lower SFR in dark green and the higher SFR in brown. The galaxies marked with a star are called outliers in the following.}
    \label{fig:rpah_scales}
\end{figure*}

\begin{figure*}[!htbp]
    \centering
    \includegraphics[width=\textwidth]{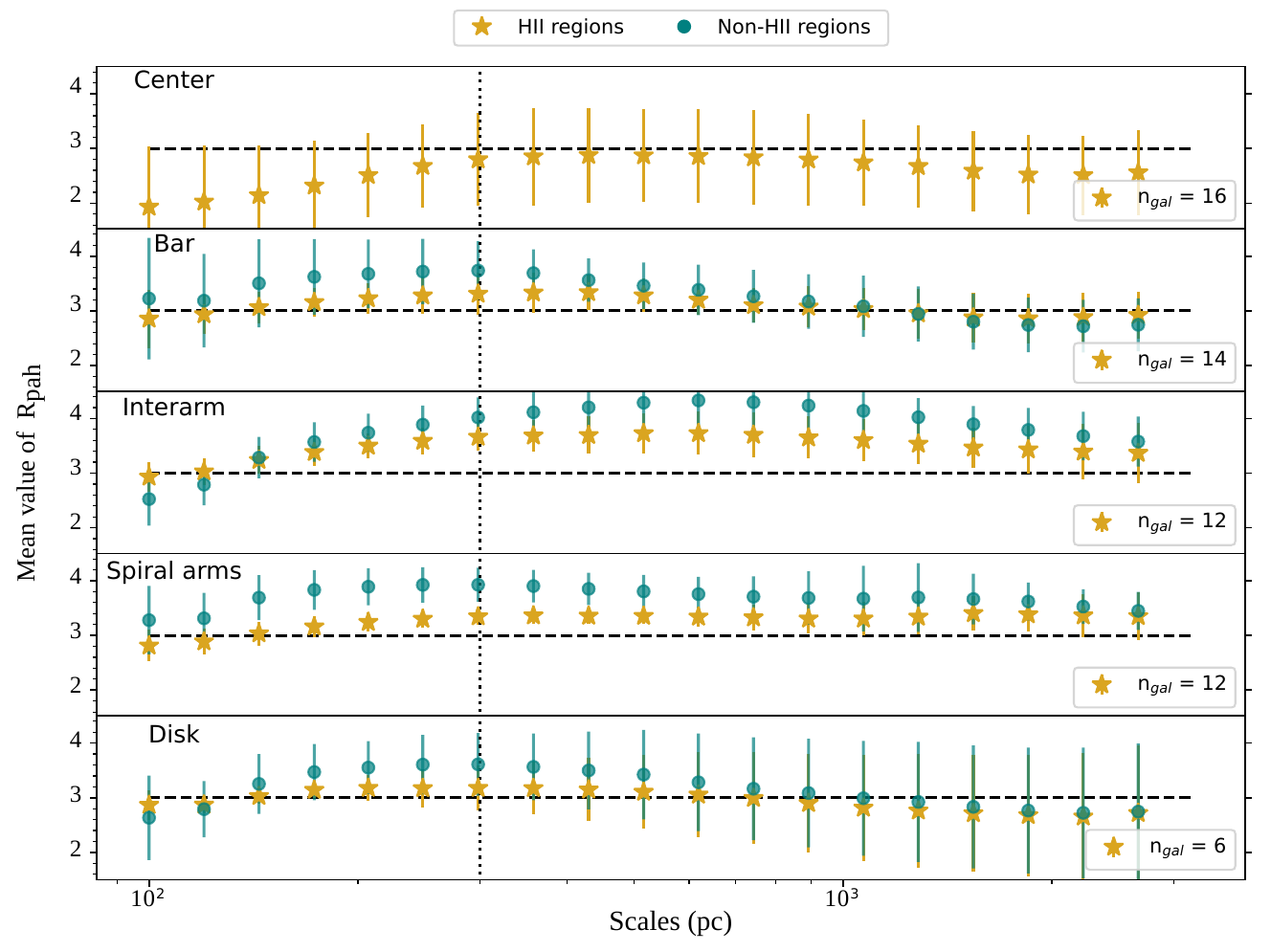}
    \caption{Mean of \Rpah{} values at a given scale for each galaxy in different environments. \HII{} regions are indicated by a star and diffuse regions by a circle. Each point shows the average of the median \Rpah{} value in all the galaxies, excluding NGC1365 and NGC3351, having this environment (written as n$_{\mathrm{gal}}$ in each plots) and the error bar presents the standard deviation of the \Rpah{} value at each scale for the whole sample. The dashed line shows \Rpah{}=$3.0$ corresponding to a PAH fraction of $4.5\%$ to $5\%$ depending on the model parameters (Figure~7 of \citealt{sutter_fraction_2023}). The dotted line correspond to the transition scale at $300$~pc.}
    \label{fig:rpah_morph_comp}
\end{figure*}

To determine whether \Rpah{} varies on different scales and to understand its correlation with environment and the presence of \HII{} regions, we show the median values of \Rpah{} for our full sample (see Fig.~\ref{fig:rpah_scales}) and in individual characteristic environments (see Figure~\ref{fig:rpah_morph_comp}).

In Figure~\ref{fig:rpah_scales}, we present the median value of \Rpah{} at each scale for each galaxy, colour-coded by the star formation rate.

We observe that \Rpah{} varies with the scale. While \Rpah{} is low at small scales, it gradually grows until reaching a maximum around $200$-$300$~pc before it stabilises at larger scales. This variation emphasizes the change of physical properties with the change of structure size. In the rest of this analysis, we will call ``transition scale'' the transition between increasing values and stable values of \Rpah{}.

We also conducted a visual inspection on the galaxies having odd behaviour compared to the rest of the sample (IC5332, NGC1365, NGC2835, and NGC3351) to determine the possible cause for their large (or small) \Rpah{} value at large scale. For NGC1365, the presence of a bright Active Galactic Nuclei (AGN) in the centre creates diffraction spikes that lead to artifacts when dividing images. For NGC3351, in addition to the diffraction spikes induced by the bright centre, the gap between the central and the outer rings creates some artefact in the decomposition at large scale, leading to lower values of \Rpah{}. Those two are excluded of our sample for the rest of our analysis.

IC5332 seems to have lower values of \Rpah{} (see \citealt{sutter_fraction_2023}) probably linked to a lower metallicity and a lower SFR. NGC2835 does not exhibit any particular features that could lead to a different behaviour in the \Rpah{} value. Furthermore, we observe in Figure~\ref{fig:rpah_scales} a small trend in SFR, where galaxies with a lower SFR tend to have lower values of \Rpah{} than galaxies with higher SFR.

We also study the impact of environment on the \Rpah{} value. We expect some dependence on environment as \HII{} regions tend to be organised along spiral arms, or as gas densities and different ISM conditions can have an impact on the PAH formation/destruction processes \citep{Matsumoto_2024}.

To do so, we compute the median value of \Rpah{} in the different environments. We separate the values from \HII{} regions and non-\HII{} regions by applying the MUSE \HII{} masks on each scale of the \Rpah{} map. We then look at the median value of \Rpah{} for \HII{} and non-\HII{} regions in each environment. The mean values for every galaxy at each scale are then shown in Figure~\ref{fig:rpah_morph_comp}. We observe that in \HII{} regions, \Rpah{} is on average lower than outside \HII{} regions. This is in agreement with \cite{sutter_fraction_2023} and \cite{Egorov_2025} who also find low \Rpah{} in HII regions. Furthermore, \Rpah{} changes with the scale in every environmental region we define (e.g. bars, arms, ...), and shows a similar trend as the one observed in Figure~\ref{fig:rpah_scales}. However, although the scale for which we observe the \Rpah{} maximum changes in different environments, it does not seem to be affected by \HII{} regions.

At large scales, both \HII{} and non-\HII{} regions show similar values. As the largest scales trace large structure and more diffuse emission, it is expected to not find any difference in the \Rpah{} values, as this large emission is more homogenous than small-scale emission. However, \Rpah{} shows different behaviours in the different environments. In the centres, we find low \Rpah{} values, below 3, which do not occur in other regions. The centres tracing mostly star-forming regions, we expect them to have lower values than the other regions, as the radiation field is high and \Rpah{} is low (see Section~\ref{sec:disc_rpah}, Pathak et al. \textit{in prep}). By looking at the variability of \Rpah{} at different scales, i.e. $R_{\mathrm{PAH}}^{j+1}-R_{\mathrm{PAH}}^{j}$, we can also identify a transition scale of the order of 200-400~pc,after which \Rpah{} values get more stable (see Appendix \ref{app:rpah_stab}, Figure \ref{fig:rpah_stab}). We also find that both the spiral arms and interarm regions exhibit higher \Rpah{} values than the rest of the regions, probably linked to the low surface density of \HII{} regions in this environment \citep[Figure 8 of][]{groves_phangs-muse_2023} and therefore fewer regions of PAH destruction. 

\section{Constraining Diffuse and Non-Diffuse Emission}
\label{PDF}
\subsection{Building the emission PDF}
\begin{figure}[!htbp]
    \centering
    \includegraphics[width=\columnwidth]{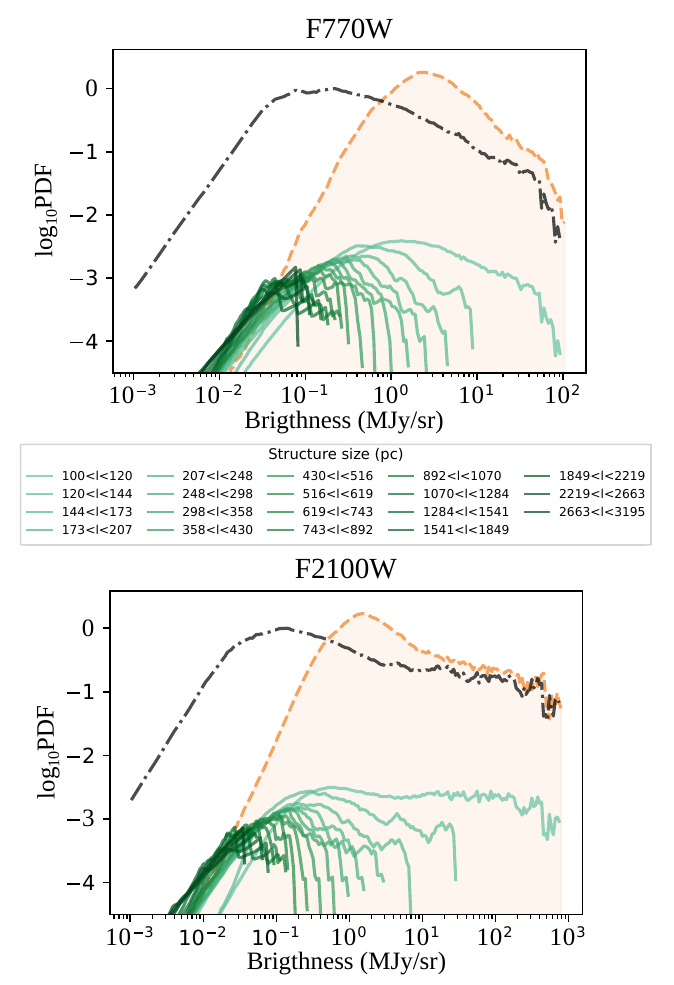}
    \caption{PDF in two different filters for NGC0628.The green lines show the PDF for each scale of the decomposition, and the black dashed-dotted line corresponds to the sum of all these PDFs. The orange, shaded PDF is the one of the original image.}
    \label{fig:pdf_gen}
\end{figure}

We derive the intensity PDF at each scale of our decomposition to understand how the emission is distributed on each scale. To do so, we compute the PDFs of the inclination-corrected intensities in each pixel in MJy~sr$^{-1}$ as done by \cite{pathak_two-component_2023}. We derive the PDF for logarithmic bins of intensity and normalise by the total intensity of the image (Eq.~\ref{eq_PDF}). We have:  
\begin{equation}
  \mathrm{PDF}(I_{\lambda_k}) = \frac{1}{\sum I_{\lambda_k}\cos i}\sum_{I_{\lambda_k}- \delta I_{\lambda}/2}^{I_{\lambda_k}+ \delta I_{\lambda}/2}I_{\lambda}\cos i,
  \label{eq_PDF}
\end{equation}
where $i$ is the inclination of the galaxy (see Table~\ref{tab:MUSE_carac}) and $\delta I_\lambda~=~0.025$~dex. The normalisation is done so that we have $\sum_k I_{\lambda_k} = 1$. We analyse the behaviour of the PDF at each scale (Fig.~\ref{fig:pdf_gen}).

When the scale increases, the flux fraction decreases (Figure~\ref{fig:pow_spec}). Furthermore, we see that larger scales trace fainter regions. By masking the centre of the galaxies and re-computing the PDF, we notice that the small scales trace mostly the centre of the galaxies, corresponding also to the part of the PDF with the higher fluxes \citep{pathak_two-component_2023}. For the rest of the analysis, we mask the centre of the galaxy using \cite{querejeta_stellar_2021} classification. Those masks are applied after the decomposition. We note that, while the structure of the PDF is conserved, the sum of the scales' PDF is not equal to the whole image PDF. This is expected, as the PDF in each scale only traces the flux distribution in the specific scale, and all PDFs are normalised by the total flux in the image. However, it means that we should be careful when analysing individual scales PDFs, and ensure that we sum the corresponding scales together before deriving the PDF.

\subsection{Distinction between small and large scales}
As described by \cite{Hennebelle_2012} and \cite{pathak_two-component_2023}, the PDF shows two regimes: a power-law regime at high fluxes and high densities, and a log-normal for lower fluxes and densities. However, the fluxes from dense regions are ``polluted'' by the IR cirrus and emission from outside of these regions. With the decomposition, we could estimate the contribution of the diffuse emission in dense regions and remove it from the total flux, which could be useful to derive an estimator purely using small-scale emission (see Section \ref{sec:sfr_est}). We investigate how the decomposition can help us to define a scale marking the separation scale between diffuse and non-diffuse emission. To identify this separation, we want to characterize which scales trace better the diffuse emission in the ISM. We sum all the images above a scale $S$ and compute the PDF of this image. We then compare this PDF with the PDF from non-\HII{} regions from the original image and iterate over $S$ to find the best scale for which the two PDFs are the closest. To do so, we use the Earth Mover Distance (EMD), a method used to describe the dissimilarity between two distributions, often naively described as the most efficient way to transform a pile of dirt to another one of equal mass, taking into account the amount of dirt moved multiplied by the distance over which it is moved \citep{Rubner_1998}. Our goal is to find the separation scale for which the EMD is the smallest between the large-scale PDF and the one from non-\HII{} regions.  

We define the EMD as:
\begin{equation}
    \mathrm{EMD}(P, Q) = \min_{\pi \in \Pi(P,Q)}\int |x-y|d\pi(x,y),
\end{equation}
where $P$ and $Q$ are the two distributions and $\Pi(P,Q)$ is the support of the joint distribution derived from $P$ and $Q$.

It is important to note that the EMD depends on the normalisation beforehand, as it assumes that both distributions have a ``mass'' of 1.

\begin{figure}[!htbp]
    \centering
    \includegraphics[width=\columnwidth]{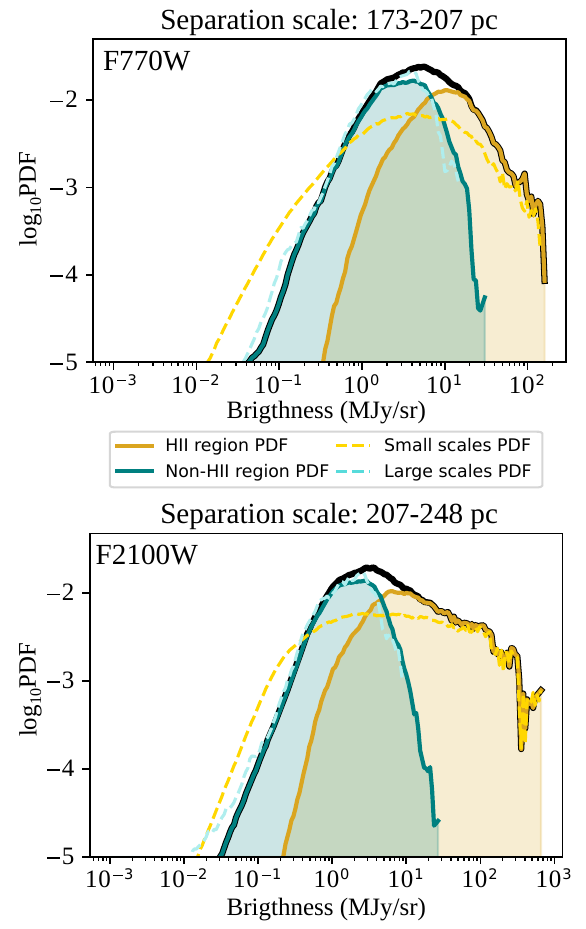}
    \caption{Best scale value in two of the four filters for NGC4252. The PDF of the original image is shown in black, the filled-gold curve is the PDF of \HII{} regions. The filled dark blue curve is the non-\HII{} regions' PDF. In dashed blue is plotted the best approximation for non-\HII{} PDF (derived using the EMD method). In yellow is shown the small-scale PDF.}
    \label{fig:pdf_best_cut}
\end{figure}

Applying the described method to the 19 galaxies analysed in this study, we find a mean separation scale between 173 and 207~pc, meaning that there is a scale, for which we can infer the emission to originate mostly from non-\HII{} regions, and that it is possible to analyse separately the emission from \HII{} regions and non-\HII{} regions. In Appendix \ref{sec:cut_val}, we present all the optimal separation scales found using the EMD method and the mean value of the separation in each filter (see Table~\ref{tab:cut_value}). We also present the comparison between the fit of the original image and the fit of the ``best-separation'' PDFs. Figure \ref{fig:pdf_best_cut} shows the best large-scale PDF for NGC0628. The dashed lines show the PDF of the summed image for scales above and below the separation scale in blue and yellow, respectively. The shaded areas show the PDF for non-\HII{} and \HII{} regions in dark blue and gold, respectively.

The shift between the small scales and the \HII{} region PDF shows how much emission in the \HII{} regions of the original image is due to emission from outside of those regions. When subtracting the difference between the maximum of the large-scale PDF and of the \HII{} region PDF, we do find a flux range similar to the one of the small-scale PDF, meaning that our small-scale image is a good proxy for the emission from dense regions only. 


\section{Discussion}\label{discussion}
\subsection{Using the decomposition}
The decomposition algorithm allows us to trace the different scales of the emission and to study them. By changing the scale input in parsec, we are able to analyse morphological properties of the emission for a large sample in different filters. However, we can argue that for the largest scales, some of the properties of the galaxies may affect the decomposition and ``contaminate'' it. The presence of a bright nucleus can lead to saturation, mostly in F2100W, and diffraction spikes. Those artifacts will not be removed by the decomposition and might affect the \Rpah{} maps. Countering this effect would be difficult, as adding masks to those regions would lead to information losses for those regions. Therefore, we need to take this phenomenon into account when looking at large scales.

As detailed in \cite{li_multiscale_2022}, each maps of the decomposition should be independent from the others. Therefore, changing the number of scale or the spacing of those scales should not change our results. With more or less scales, each scale would have a lower or higher flux fraction respectively, but this would only change the cut value (i.e. the scale number) and not the physical scale of the transition as we are summing images before deriving the PDF. 

\subsection{Using PDF shape to identify diffuse emission}

The PDF provides information on the way the emission is distributed amongst the different scales, even though some of the features are different when looking in different filters. We observe that the power-law component is mostly visible in the 21\um{} band and less present in the other three bands (see Figure~\ref{fig:pdf_gen}). This is in agreement with \cite{pathak_two-component_2023} who used the same sample, but also with Milky Way studies such as \cite{Kainulainen_2014}. The 21\um{} band mostly traces the dust continuum, whereas the 7.7 and 11.3\um{} bands trace PAHs and 10\um{} traces small grains. Those grains (PAHs and small grains) are often destroyed in environments with high radiation, such as \HII{} regions \citep[][and Section~\ref{sec:disc_rpah}]{Chastenet_2023a, Egorov_2023}, and the remaining dust has an increased thermal emissivity for wavelengths longer than 20\um{} \citep{draine_excitation_2011}. While PAH destruction decreases MIR emission in \HII{} regions, the increased thermal emissivity enhances emission in \HII{} regions in the 21\um{} band, leading to an extended power-law component in this band compared to the others. Furthermore, when looking at the high-intensity end of the PDFs (especially in F2100W), we see that the PDF of the first scale traces well the original one, as it is mostly dense and small regions which contribute to this part of the PDF. 

As seen in Figure~\ref{fig:pdf_gen}, the CDD induce a shift in flux when computing the individual PDFs. To account for this, we sum different scales together and then derive the PDFs (Figure~\ref{fig:pdf_best_cut}). To verify our method and the value of the transition scale we found, we compare the fits of our PDFs (a detailed analysis is shown in Appendix~\ref{sec:cut_val}). Overall, we find good agreements when comparing the log-normal fits, with a average difference of $\Delta\mu~=~0.08\pm0.12$ for all four filters (see Figure~\ref{fig:diff_fit_EMD}). When comparing the power-law fits, we find accurate fits mostly for the F2100W filter, as this component of the PDF is mostly visible in this band. It ensures that the transition scale we found using diffuse emission also allows to isolate dense emission. 

\subsection{Impact on Star Formation Rate estimators}\label{sec:sfr_est}
The SFR is a parameter allowing us to trace galaxy evolution and the efficiency of conversion of the gaseous component to stars. However, measuring the SFR with high accuracy can prove challenging due to dust obscuration. Furthermore, to build relevant SFR estimators, we have to ensure that the luminosities and the emission we use to derive the SFR trace young stellar populations \citep[under $200$~Myr; see e.g.][]{kennicutt_star_2012}. The young and massive stars ionizing their surroundings offer good tracers as they are short-lived and can be associated with \HII{} regions. Emission outside of those regions is often associated with older stellar populations that are not relevant to forming coherent SFR estimators and therefore need to be controlled and removed to derive reliable estimators. To build SFR estimators, one can use different tracers such as UV, nebular, or dust emission.

The UV emission originates directly from the photosphere of massive young stars and is therefore the main window to young star formation. However, it is highly attenuated by dust \citep[e.g.][]{Burgarella_2013}. H${\alpha}$, an optical tracer of recombination in \HII{} regions, is also often used as it is very intense and is less attenuated than bluer lines. However, it only traces unobscured star formation. Furthermore, UV is sensitive to stellar population up to $100$~Myr, while H${\alpha}$ is sensitive to more ``instantaneous'' star formation, up to $10$~Myr \citep{kennicutt_star_2012}. By using IR estimators, we measure light from young hot stars absorbed and re-emitted by the dust. For example, in dusty starburst galaxies, the SFR is proportional to dust bolometric luminosity \citep{calzetti_calibration_2007}. However, as seen before, IR fluxes are also contaminated by dust heated by active nuclei, older stellar populations, and non-star-forming regions, which bias the estimators \citep{belfiore_calibration_2023, Gregg_2025}. Therefore, it is important to calibrate the different SFR estimators for their respective flaws to build a robust method to derive SFR.

\begin{figure}[!hbtp]
    \centering    
    \includegraphics[width=\columnwidth]{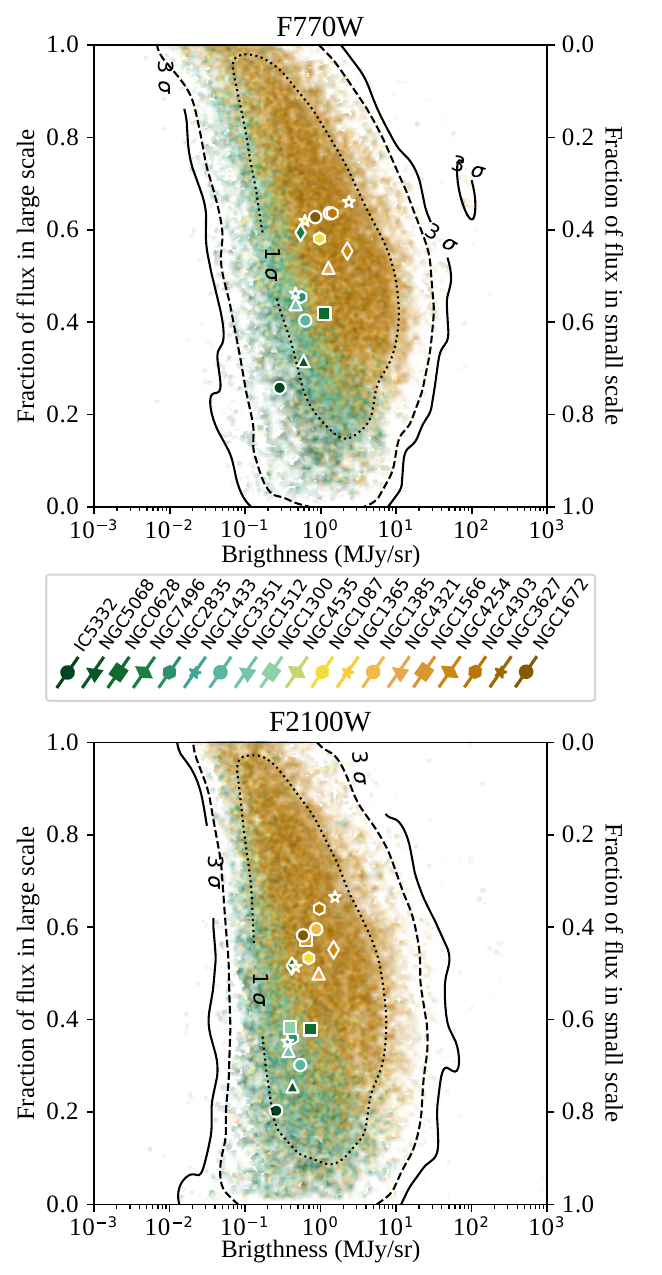}
    \caption{Fraction of flux inside of \HII{} regions in large-scale images compared to \HII{} regions flux in small-scale images. Each point is the mean flux fraction in the large scale image as a function the mean brightness in the small scale image for an \HII{} region. Galaxies are colour-coded by SFR in increasing order. The black contours show the $1\sigma$ (dotted), $2\sigma$ (dashed), $3\sigma$ (plain line) of the total distribution. The coloured markers highlight the median values for each galaxies.}
    \label{fig:frac_flux_all}
\end{figure}

Using the decomposition, we are able to identify the emission originating from dense regions and separate it from the more diffuse and large scale emission, originating from dust heated by older and more diffuse stellar population. In Figure~\ref{fig:frac_flux_all} we present the fraction of flux inside \HII{} regions in large-scale images (above $207$~pc) compared to the flux of small-scale emission in the same regions.
The distribution is similar in different filters (here are shown F770W and F2100W only), but the flux fraction does vary with SFR. Galaxies with higher SFR tend to have higher fluxes at small scales in \HII{} regions but also a higher fraction of flux in the large-scale images. We also observe a bimodal distribution with SFR. This distribution correlates with the distance of the main-sequence of star forming galaxies. Galaxies with a positive distance relative to the main sequence of galaxies have higher diffuse fraction whereas galaxies with a negative distance to the main sequence have a lower flux fraction (see Figure 13 of \citealt{Maschmann_2024}). This feature could exhibit the role of feedback from stellar evolution in the dispersion of gas and dust leading to higher diffuse fraction of the emission even in dense structures. The higher fraction of diffuse flux for higher SFR could show a contamination of IR estimators by large-scale and diffuse emission.

\subsection{Destruction of PAH in \texorpdfstring{\HII{}}{} regions and at small scales} \label{sec:disc_rpah}
We observe that \HII{} regions have lower \Rpah{}. This is in agreement with previous studies in the Milky Way and Magellanic Clouds \citep{Sandstrom_2010, Croiset_2016} but also with extragalactic studies \citep{Relano_2018,Egorov_2023, sutter_fraction_2023, Egorov_2025}. We also find a correlation between SFR and \Rpah{} in agreement with extragalactic studies \citep{Shipley_2016, Figueira_2022} where the PAH emission is used as a tracer for SFR (Figure~\ref{fig:rpah_scales}). We looked at the impact of metallicity, stellar mass, and specific SFR, but no clear trends were seen. As our galaxies have high metallicities but only cover a narrow dynamical range ($8.3 < 12+\log({\rm O/H})<8.6$), it is difficult to judge the existence of a trend of the PAH fraction with metallicity in our sample (refer to \citealt{Egorov_2025} for metallicity trends).

Furthermore, we find that smaller scales have lower \Rpah{} up to a scale of around $200$~pc. Linking scales with the nebular regions, we do expect lower \Rpah{} for the small scales as they trace smaller and denser structures such as \HII{} regions. The behaviour of \Rpah{} at small scales and in \HII{} regions suggests that PAH are destroyed in those regions and that PAHs lifetime is smaller than \HII{} regions \citep{Kim_2025}. Several destruction mechanisms and regulation processes have been proposed in the literature. The radiation field, its intensity, and its hardness \cite[the ratio of the fluxes over hard and soft UV ranges as defined by][]{Egorov_2023}, could influence PAH destruction. Some theoretical works also propose that PAHs can be sputtered and fragmented due to atomic and electronic interactions, leading to their destruction. Previous studies show lower PAH emission in galactic \HII{} regions \citep{Povich_2007} and a link between a lower PAH fraction and a higher intensity of the radiation field.

\cite{Lind_power_2025} describe the existence of a break scale $l_0 = 160 ^{+110}_{-50}$~pc in the power spectrum of PAH sensitive bands. They show that below this characteristic scale, PAH emission seems to be suppressed, which would be consistent with PAH destruction in \HII{} regions and with the idea that this scale could distinguish the dense ISM from the diffuse ISM. Comparing with our study, we find a similar scale using the PDFs to trace the origin of the emission, and a slightly larger scale when focusing on \Rpah{}.

As shown by the PDFs (see Section~\ref{PDF}), above $200$~pc, we mostly have emission from diffuse regions. Having a higher value of \Rpah{} could indicate that PAHs are less destroyed in those environments than in dense regions, leading to a higher PAH fraction. This informs us about the nature of this emission, originating from larger structures, richer in PAHs, and with lower radiation intensities. When looking at large scales (above $500$~pc), we trace very diffuse emission, where the radiation field is probably lower and where the stochastic heating of the grains might dominate. The transition scale found using the CDD could be an indicator of the scale at which destructive processes come into equilibrium with formation processes, leading to more constant values at larger scales where PAH are no longer destroyed.

\subsection{Flux-weighted \texorpdfstring{\Rpah{}}{}}\label{fluxwei_discussion}
Following the appendix in \cite{sutter_fraction_2023}, we compare the pixel-weighted average, where each pixel in the \Rpah{} maps has the same weight, to the flux-weighted average, where the highest fluxes have the highest weights, and study the new \Rpah{} at each scale.

\begin{figure*}[!htbp]
    \centering
    \includegraphics[width=\textwidth]{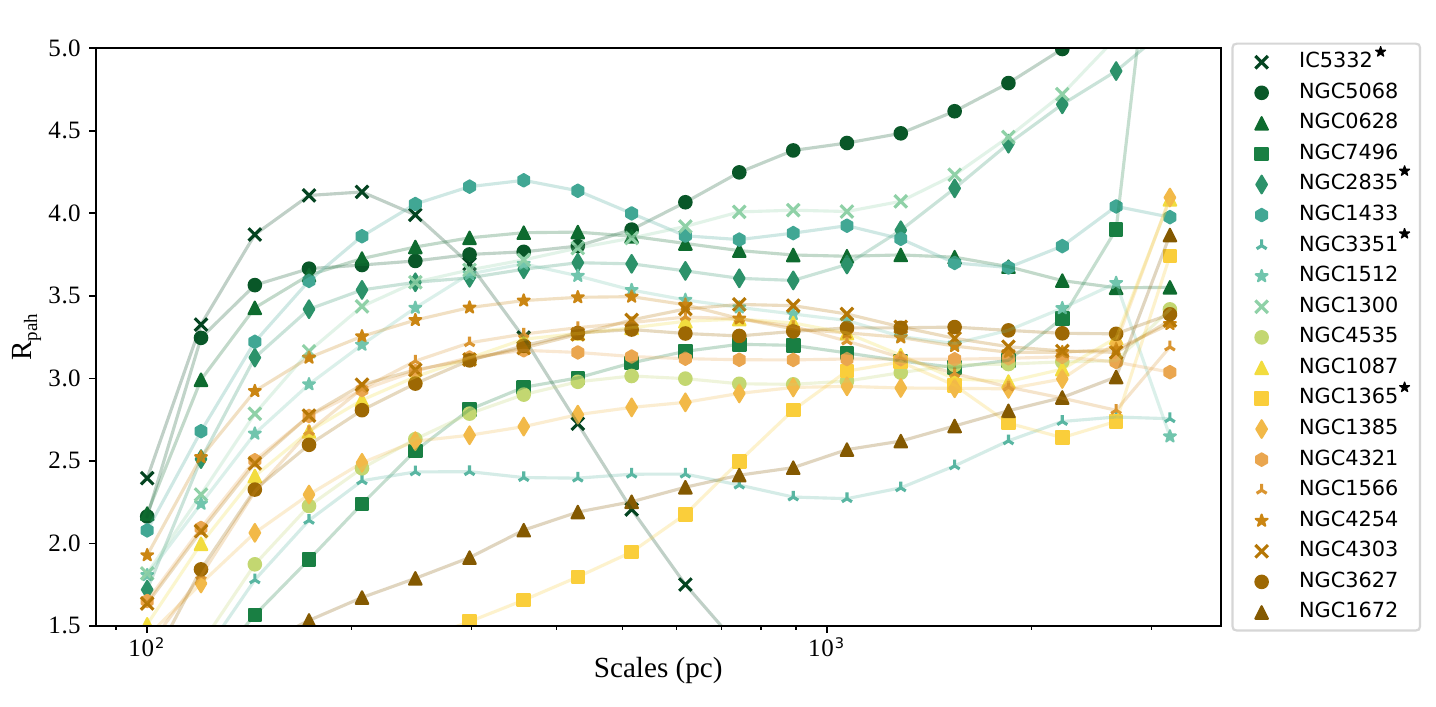}
    \caption{Flux weighted \Rpah{} for all galaxies at each scale. Galaxies are colour-coded by SFR in increasing order. The outliers of Figure~\ref{fig:rpah_scales} are marked with a star.}
    \label{rpah_flux_weight}
\end{figure*}

We find in Figure~\ref{rpah_flux_weight} that flux-weighted values of \Rpah{} are lower, as also found by \cite{sutter_fraction_2023}.
However, even though there are more outliers, specifically on large scales, we see that \Rpah{} follows the same trend as the pixel-weighted ones, with low values on small scales, an increase until $300$~pc, and a stabilisation on large scales. Furthermore, the outliers defined in Section~\ref{sec:rpah_med} that had on average higher values of \Rpah{} tend to have here lower \Rpah{}. Looking at flux-weighted values reduces non-physical values created by the division of the fluxes, which could lead to strong edges in some of the maps.

When looking at nebular regions and the environmental repartition of flux-weighted \Rpah{} (see Appendix~\ref{app:rpah_flux_morph}, Figure~\ref{fig:rpah_flux_morph}), we find that \HII{} regions have lower values than for the pixel-weighted average but that non-\HII{} regions tend to have slightly higher values. However, the global trend seen using the pixel-weighted median is also seen for flux-weighted medians. 

This phenomenon can be explained by \cite{pathak_two-component_2023} and by our study of the emission PDF on different scales. We observe higher fluxes in F2100W for small scales than in the other filters (see Figure~\ref{fig:pdf_gen} and Appendix~\ref{app:pow_spec}). These bright regions dominate in the total F2100W values and therefore lower the value of \Rpah{} on small scales. Furthermore, these bright regions correspond to \HII{} regions, lowering \Rpah{} when looking at those regions. However, outside of \HII{} regions, we notice the opposite trends, where F770W and F1130W have larger contributions than F2100W, leading to higher values of \Rpah{}.


\section{Conclusion}\label{conclusion}

To understand the way star-forming regions influence their surroundings and the interplay between dense \HII{} regions and the larger scale of the galaxy, it is important to trace the different phases of the ISM across varying physical scales.

In this work, we present an analysis of emission scales using the Constrained Diffusion Method on JWST mid-IR maps in 19 nearby galaxies from the PHANGS sample. Following previous work, we bring a new methodology to understand the way the different structures contribute to the emission and their characteristics by studying both the PAH fraction and the shape of the PDF on different scales. We adapted the Constrained Diffusion algorithm to our specific case, made it more flexible on the input scale choice, and used it on a different set of scales to probe efficiently the multi-scale structure of galaxies. Using this decomposition, we analysed different properties of dust and its emission in environmental and nebular regions of galaxies.
\begin{itemize}
    \item We measure \Rpah{} at varying scale using F770W, F1130W, and F2100W. Using our decomposition, we study the evolution of \Rpah{} at each scale, in different galactic and physical environments. We find that \Rpah{} is scale-dependent and that it increases until it reaches a stable-value around $200$ to $400$~pc. This could indicate that the PAHs are destroyed by \HII{} regions even beyond their limits. We also have an overview of the distribution of \Rpah{} in different environments, and we confirm that while we observe the same general trends, \HII{}-regions exhibit lower \Rpah{}. \\

    \item Using the intensity PDF, we ensured that our decomposition kept the expected distribution, even though it induced a shift in the flux. We also used our decomposition to characterize a transition scale between the non-\HII{} regions, which have a log-normal PDF, and the \HII{} regions that have a power-law PDF. We find a cut value between $173$ and $207$~pc using the Earth Mover Distance, which could be an estimate of the scale at which we can separate diffuse emission of dust heated by surrounding stars from the emission directly coming from the UV light reprocessed by dust.\\

    \item We highlight the destruction of PAHs in strong-radiation environments and find a transition scale of $\sim200$~pc which may correspond to the characteristic size where \HII{} region feedback transitions to ISM heating. Using the decomposition, we could remove the large-scale emission and therefore isolate the emission of young star-forming regions and dense regions. This method using the Constrained Diffusion algorithm could be an helpful tool to derive SFR estimators and maps of dense regions.
\end{itemize}
\begin{acknowledgements}
We thank the anonymous reviewer for their comments, advices and suggestions which contributed to improve this paper. MB acknowledges support by the ANID BASAL project FB210003. This work was supported by the French government through the France 2030 investment plan managed by the National Research Agency (ANR), as part of the Initiative of Excellence of Université Côte d’Azur under reference No. ANR-15-IDEX-01. This research was funded, in whole or in part, by the French National Research Agency (ANR), grant ANR-24-CE92-0044 (project STARCLUSTERS). We thank the German Science Foundation DFG for financial support in the project STARCLUSTERS (funding ID KL 1358/22-1 and SCHI 536/13-1). OE acknowledges funding from the Deutsche Forschungsgemeinschaft (DFG, German Research Foundation) -- project-ID 541068876. The National Radio Astronomy Observatory and Green Bank Observatory are facilities of the U.S. National Science Foundation operated under cooperative agreement by Associated Universities, Inc. HAP acknowledges support from the National Science and Technology Council of Taiwan under grant 113-2112-M-032-014-MY3. \\

\textit{Software:} \texttt{astropy} \citep{astropy:2022}, \texttt{matplotlib} \citep{Hunter:2007}, \texttt{numpy} \citep{harris2020array}, \texttt{scipy} \citep{2020SciPy-NMeth}, \texttt{reproject}, SAOImageDS9 \citep{saods9}
\end{acknowledgements}
 

\bibliographystyle{aa.bst}
\bibliography{biblio.bib}


\begin{appendix}
\section{Constrained Diffusion Algorithm and scale spectra}
\subsection{Description of the Constrained Diffusion Algorithm}\label{app:dec_full}

\begin{figure*}[h!]
    \centering
    \includegraphics[width=\textwidth]{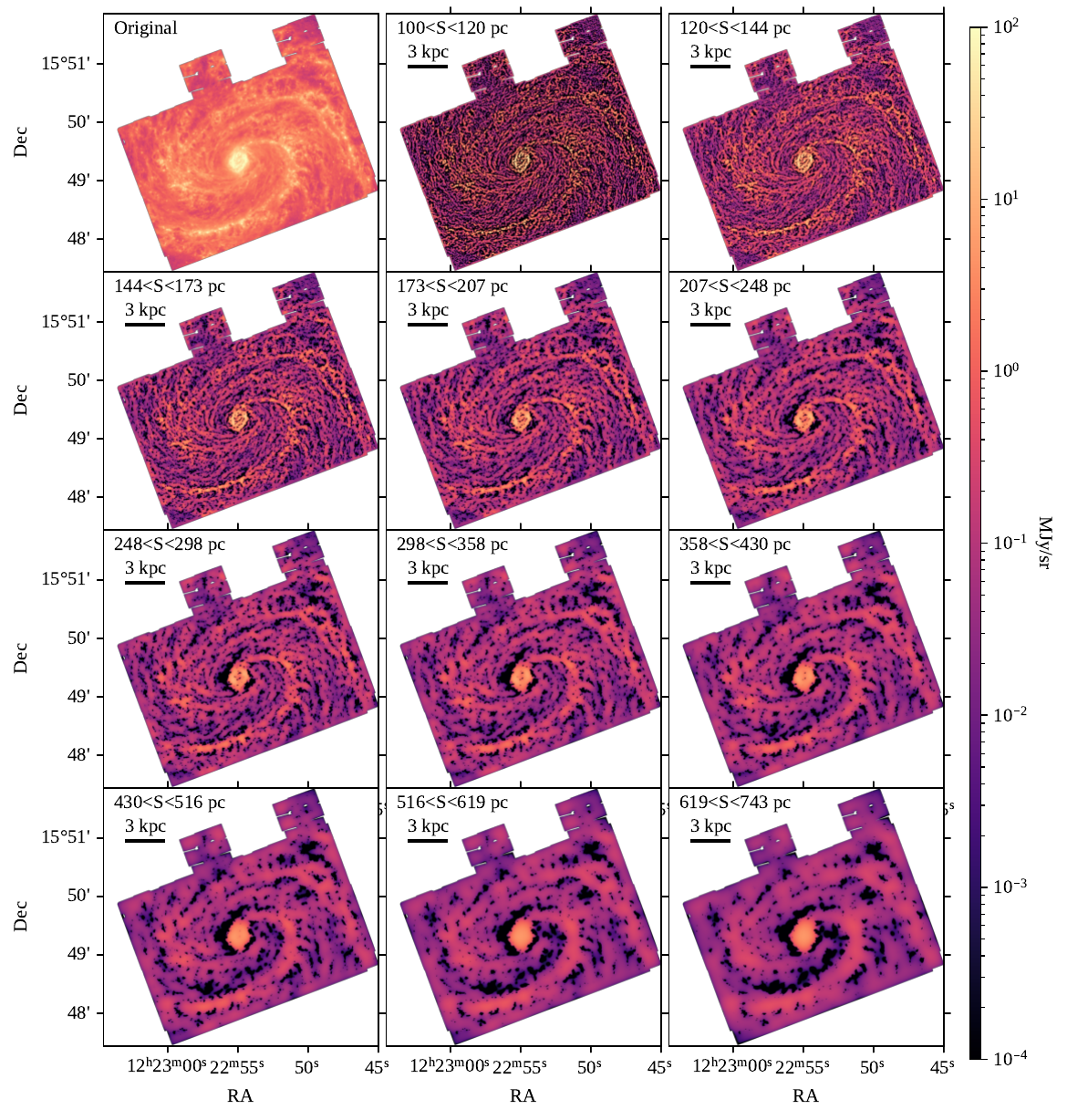}
    \caption{Decomposition on different scales for NGC4321 in 7.7\um{}.}
    \label{fig:dec_more_sc}
\end{figure*}

With the Constrained Diffusion method, the image is decomposed by solving a modified version of the diffusion equation: 
\begin{equation}
    \frac{\partial I_t}{\partial t} = \mathrm{sgn}(I_t)\mathcal{H}(-\mathrm{sgn}(I_t)\nabla^2I_t)\nabla^2I_t,
\label{equationdec}
\end{equation}
where $\mathcal{H}$ is the Heavyside function, sgn is the sign function, and $I_t$ corresponds to the intensity of the maps at each $x$ and $y$ position at an effective time $t$, and we ensure to have invariant results for $I_t \xrightarrow{}-I_t$. We consider a 2D map (as our image) of shape ($n_x$, $n_y$) and a given set of scales $S$ on which the image is going to be decomposed. The algorithm takes $I(x,y)$ as input and solves equation \ref{equationdec} with $t$ from $t=1$ to $t=S_{\rm max}^2/2$. We construct the decomposed maps $M_n = I_n - I_{n-1}$ where $M_n$ enclosed structure with a size within $S = 2^{n-1}$ and $S = 2^{n}$. We therefore have a set of images such as $I(x,y) = \sum_n M_n(x,y)$. To generate the maps, we move from $t_1$ to $t_2$ in $n$ steps, solving:
\begin{equation}
    I_{t_1+\delta t} = I^+_{t_1+\delta t}  + I^-_{t_1+\delta t},
\end{equation}
where: 
\begin{align}
    I^+_{t_1+\delta t}  &= \mathrm{min}\left(I_{t_1}*\mathcal{G}\left(\!\sqrt{2\delta t}\,\right), I_{t_1}\right), \\
    I^-_{t_1+\delta t}  &= \mathrm{max}\left(I_{t_1}\!*\mathcal{G}\left(\!\sqrt{2\delta t}\,\right), I_{t_1}\right), 
\end{align}
where $\mathcal{G}$ is the Gaussian function and $*$ is the convolution product. The component maps, containing the structures of a given size, are then obtained by subtracting the smoothed maps two-by-two. We present a version of the decomposition on NGC4321 with more scales in Figure~\ref{fig:dec_more_sc}
\begin{figure*}[h!]
    \centering
    \includegraphics[width=\textwidth]{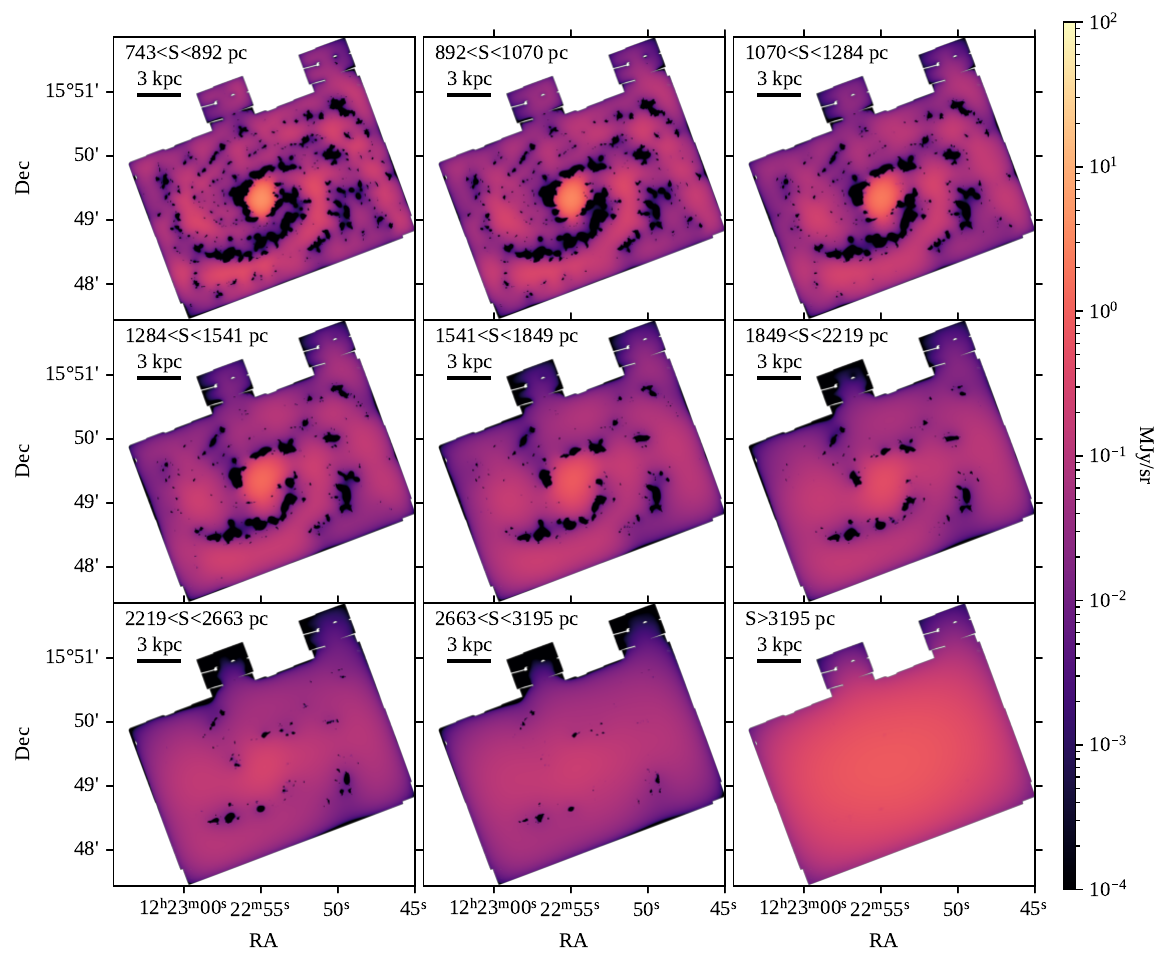}
    \caption{Figure \ref{fig:dec_more_sc} continued.}
    \label{fig:dec_more_sc2}
\end{figure*}

\subsection{Scale spectra and comparison with the Fourier Transform}\label{app:pow_spec}

\begin{figure}[!h]
    \centering
    \includegraphics[width=\columnwidth]{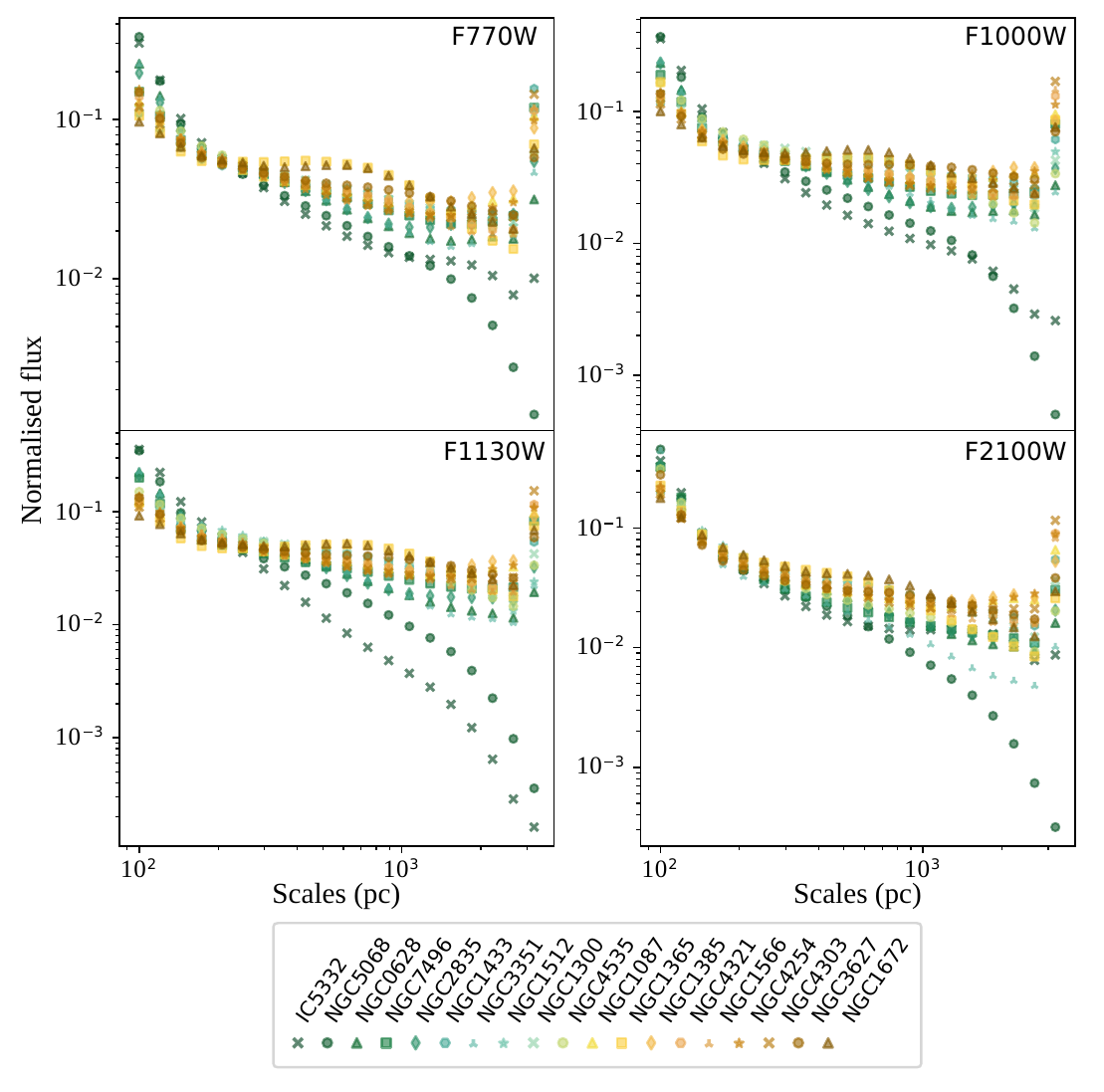}
    \caption{Normalised flux at each scale for every galaxy color-coded by SFR in the four filters.}
    \label{fig:pow_spec}
\end{figure}

    We studied the flux repartition in each scale compared to the original flux of the image (Figure~\ref{fig:pow_spec}). The scale spectrum shows the same trend for every filters with higher flux fraction for smaller scales. Furthermore, we observe the last scale having a lot of dispersion, and, on average higher values than the previous scales. This lead to more dispersed values of \Rpah{} on the last scale and is probably due to an artifact induced by the decomposition. By looking at the cumulative distribution function or the total flux, we also confirm that the flux is conserved by the decomposition. 

    Compared to a Fourier transform, the CDD isolate structures of each size without taking periodicity into account. The simplest example of this would be to look at a dirac signal, with null value everywhere except for one point. Using the CDD and computing the scale spectrum on this signal, we expect to recover all of the flux for the one pixel scale whereas the Fourier transform would give a constant power spectrum for all spatial frequencies, and therefore for all spatial scales. If we were to add more dirac, the Fourier power spectrum would change, whereas the scale spectrum will always have the same shape for the CDD as this method is not sensitive to the periodicity of structures. A other naive way to think about it would be to consider the CDD is only looking at the peaks in the signal whereas the Fourier transform diffuses small scale structures in the spatial frequency space (see the example of the dirac) and add power to periodic structures.
    
\begin{figure*}[!h]
    \centering
    \includegraphics[width=\textwidth]{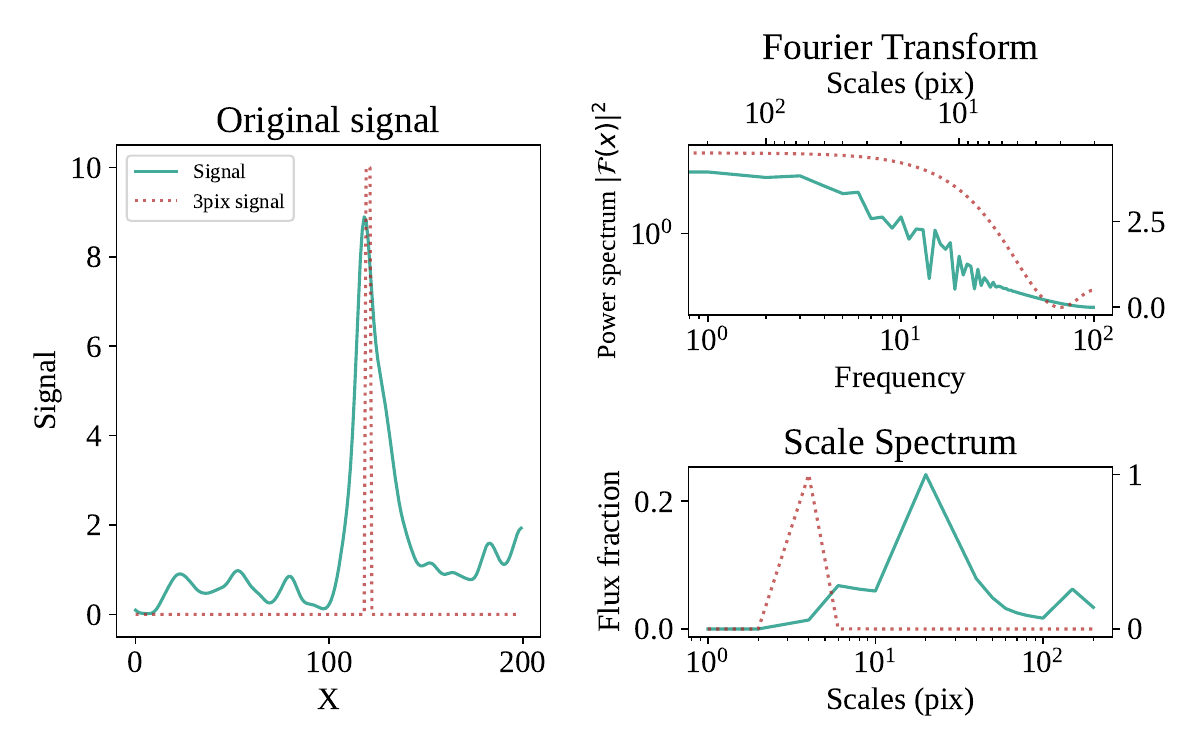}
    \caption{Comparison between the Fourier Transform and the CDD applied to a boxcar of 3 pixels (red dotted curves) and to a 1D astrophysical signal (teal curves).}
   
    \label{fig:fft_cdd}
\end{figure*}

    In Figure~\ref{fig:fft_cdd} we applied the Fourier transform (upper right plot) and the CDD (bottom right plot) to two examples of 1D signals. The first one is a simple 3-pixels-boxcar function (dotted red line, hereafter ``the 3 pixel signal''), and the second one is a 200 pixel cut in the X-axis of the F770W map of NGC0628 (solid teal lines, hereafter``astrophysical signal''). Looking at the 3 pixel signal, we recover the expected \textit{sinc} shape of the power spectra using the Fourier transform. Using the CDD we have all the flux in the structures of size between 2 and 4 pixels, corresponding to our 3 pixel signal.
    
    Looking at the astrophysical signal, we observe that the power spectrum created using the Fourier Transform is decreasing from lower spatial frequencies (higher spatial scales) to higher spatial frequencies (lower spatial scales). The scale spectrum derived using the CDD has a peak for structures between 10 and 50 pixels wide, meaning that most of the flux of the original signal is distributed around these scales. 
    
    The scale spectra is a way to identify the flux distribution in different scales, independently from the underlying periodicity of the original signal.

\section{Separation scales obtained with the EMD method}\label{sec:cut_val}

To determine the best cut value between the large and small scales, we used the Earth Mover Distance (EMD). After several tests using different methods, such as comparing averages and standard deviation, we decided to only use the EMD. 

\begin{table}[!htbp]
\begin{tiny}
\centering
    \begin{tabular}{c|cccc}
        Galaxy & F770W & F1000W & F1130W & F2100W \\
        \hline
        \hline
        IC5332 & 100--120 & 100--120 & 100--120 & 100--120\\
        \hline
        NGC0628 & 120--144 & 120--144 & 120--144 & 120--144\\
        \hline
        NGC1087 & 248--298 & 248--298 & 248--298 & 298--358 \\
        \hline
        NGC1300 & 144--173 & 120--144 & 120--144 & 120--144 \\
        \hline
        NGC1365 & 173--207 & 144--173 & 144--173 & 144--173 \\
        \hline
        NGC1385 & 619--743 & 516--619 & 516--619 & 516--619 \\
        \hline
        NGC1433 & 120--144 & 100--120 & 120--144 & 100--120 \\
        \hline
        NGC1512 & 120--144 & 100--120 & 120--144 & 100--120 \\
        \hline
        NGC1566 & 207--248 & 207--248 & 207--248 & 207--248 \\
        \hline
        NGC1672 & 207--248 & 207--248 & 207--248 & 207--248 \\
        \hline
        NGC2835 & 120--144 & 120--144 & 120--144 & 120--144 \\
        \hline
        NGC3351 & 100--120 & 100--120 & 100--120 & 100--120 \\
        \hline
        NGC3627 & 248--298 & 248--298 & 248--298 & 358--430 \\
        \hline
        NGC4254 & 173--207 & 173--207 & 173--207 & 207--248 \\
        \hline
        NGC4303 & 248--298 & 248--298 & 248--298 & 298--358 \\
        \hline
        NGC4321 & 144--173 & 144--173 & 144--173 & 144--173 \\
        \hline
        NGC4535 & 144--173 & 144--173 & 144--173 & 144--173 \\
        \hline
        NGC5068 & 120--144 & 100--120 & 100--120 & 120--144 \\
        \hline
        NGC7496 & 144--173 & 144--173 & 173--207 & 144--173 \\
        \hline
        \hline
        Mean & 169--203 &  160--192 & 165--198 & 170--205 \\
    \end{tabular}
    \caption{Optimal separation scales obtained using the Earth Mover Distance. The mean values are computed separately for the lowest and highest values and exclude NGC1365 and NGC3351.}
    
    \label{tab:cut_value}
    \end{tiny}
\end{table}

The Earth Mover Distance offers a robust alternative to other tests between distributions, such as comparing averages or the Kolmogorov-Smirnov (KS) test. When comparing distributions, one needs to define a question that will be answered by the statistical properties of the distributions. In our case, we want to find the closest distribution, meaning we want to be able to measure the gap between the distributions and select the smallest gap. The KS test only gives information on the type of distribution, and if both our distributions are drawn from the same distribution, it cannot be used to minimize the distance between distributions. By comparing averages of distributions, we might not know whether the distributions are drawn from the same type of distribution, but we can minimize the distance between averages.

In our case, we assume the distributions are both drawn from a log-normal distribution and want to find the cut value for which the two distributions are the closest. To do so, and for each galaxy, we compute the PDF of non-\HII{} region from the original image and use it as our reference PDF. We then choose a scale S, sum all the decomposed image above this scale S and derive the PDF. If the galaxy has a identified centre in \cite{querejeta_stellar_2021} masks, we mask this centre before deriving the PDFs. We use \texttt{scipy.stats.wasserstein\_distance} to compute the EMD between the original PDF of non-\HII{} regions and the large-scale PDF. We then iterate on each value of S to find the scale for which the EMD is minimal. We present those results in Table \ref{tab:cut_value}.

\begin{figure}
    \centering
    \includegraphics[width=\columnwidth]{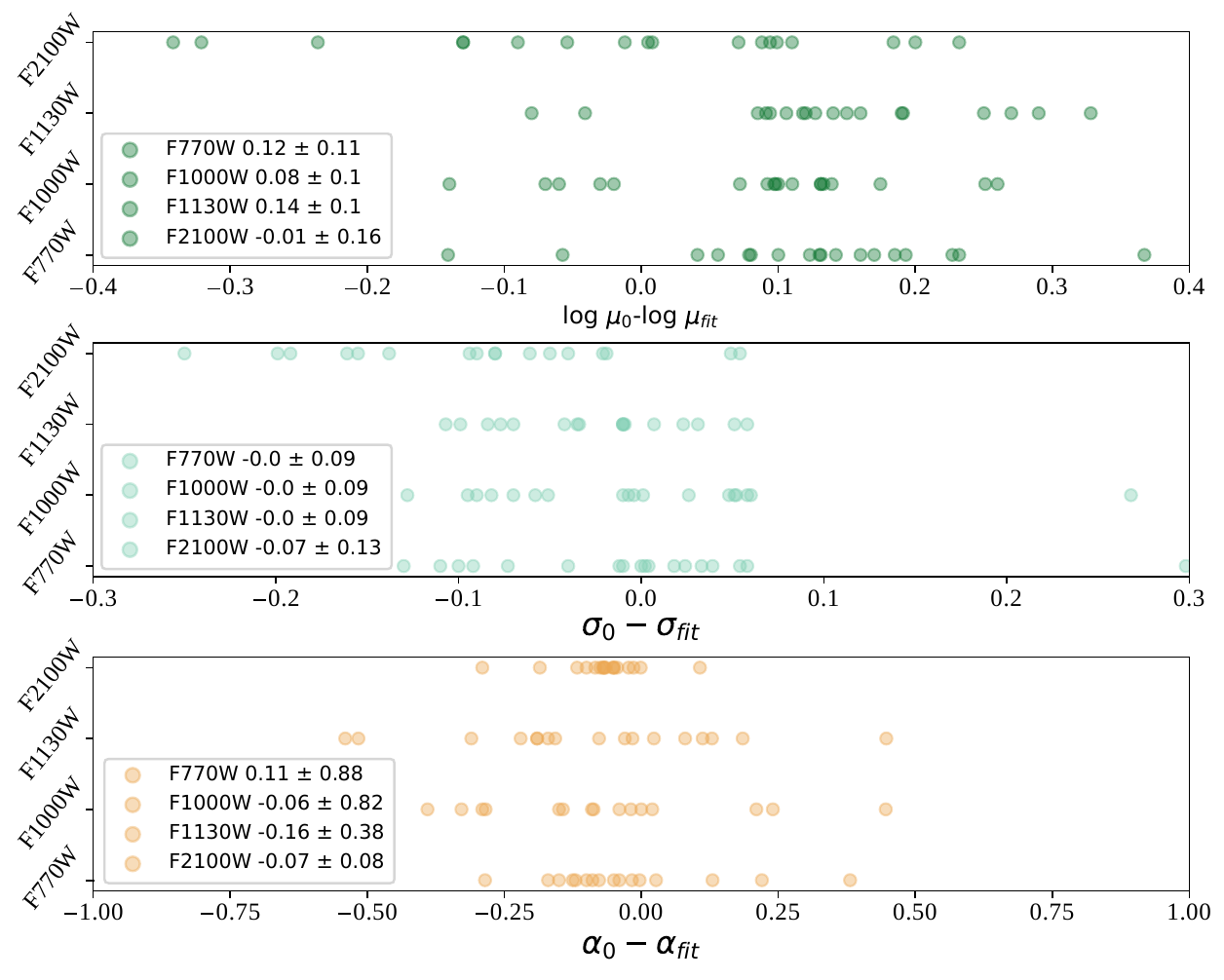}
    \caption{Difference between the best fit of \HII{} and non-\HII{} PDF and the best cut PDF. $\mu$ and $\sigma$ are the parameters of the log-normal (large-scale component), and $\alpha$ is the power law parameter. Each point represents the difference for a galaxy. The global results present the mean results and the standard deviation in each filter.}
    \label{fig:diff_fit_EMD}
\end{figure}

To check whether the separation scales found using the EMD are reliable we first perform visual inspection to compare the PDF derived at the best separation scale. We also fit both part of the PDFs with a power-law and a lognormal function and compare the parameters of the fit for the PDFs of the original image and the PDF derived at the best scale previously defined (see Figure~\ref{fig:diff_fit_EMD}).
We define the power-law PL as:
\begin{equation}
    \mathrm{PL} = bx^{\alpha},
\end{equation}
and the log-normal LN as:
\begin{equation}
    \mathrm{LN} =  A\exp(-[\log(x)-\mu)^2]/2\sigma^2),
\end{equation}
where $x$ is the flux in MJy/sr. We only look at the difference between $\alpha$, $\mu$ and $\sigma$ between the original and the EMD defined PDFs.
We find a systematic error of $0.1$~dex between the $\mu$ parameter of the non-\HII{} original PDF, and the large-scale fit in every filter but F2100W, where the scatter looks larger. However, this systematic is absent when looking at $\sigma$. We also find good agreement when looking at the parameters of the power-law fit (bottom plot of Figure~\ref{fig:diff_fit_EMD}). However, we find large scatter,  mostly in F770W and F1000W. In those filters, the power-law component is less intense, which could lead to more errors in the fitting process. Looking at F2100W, where the power-law component is more visible, we notice that the scatter is small, and that our method allows us to reconstruct the power-law component in the filters where it is the most visible.

\section{Supplemental figures of \texorpdfstring{\Rpah{}}{} statistics}

\subsection{Flux Weighted \texorpdfstring{\Rpah{}}{} in environmental regions}\label{app:rpah_flux_morph}
\begin{figure*}[!htbp]
    \centering
    \includegraphics[width=13cm]{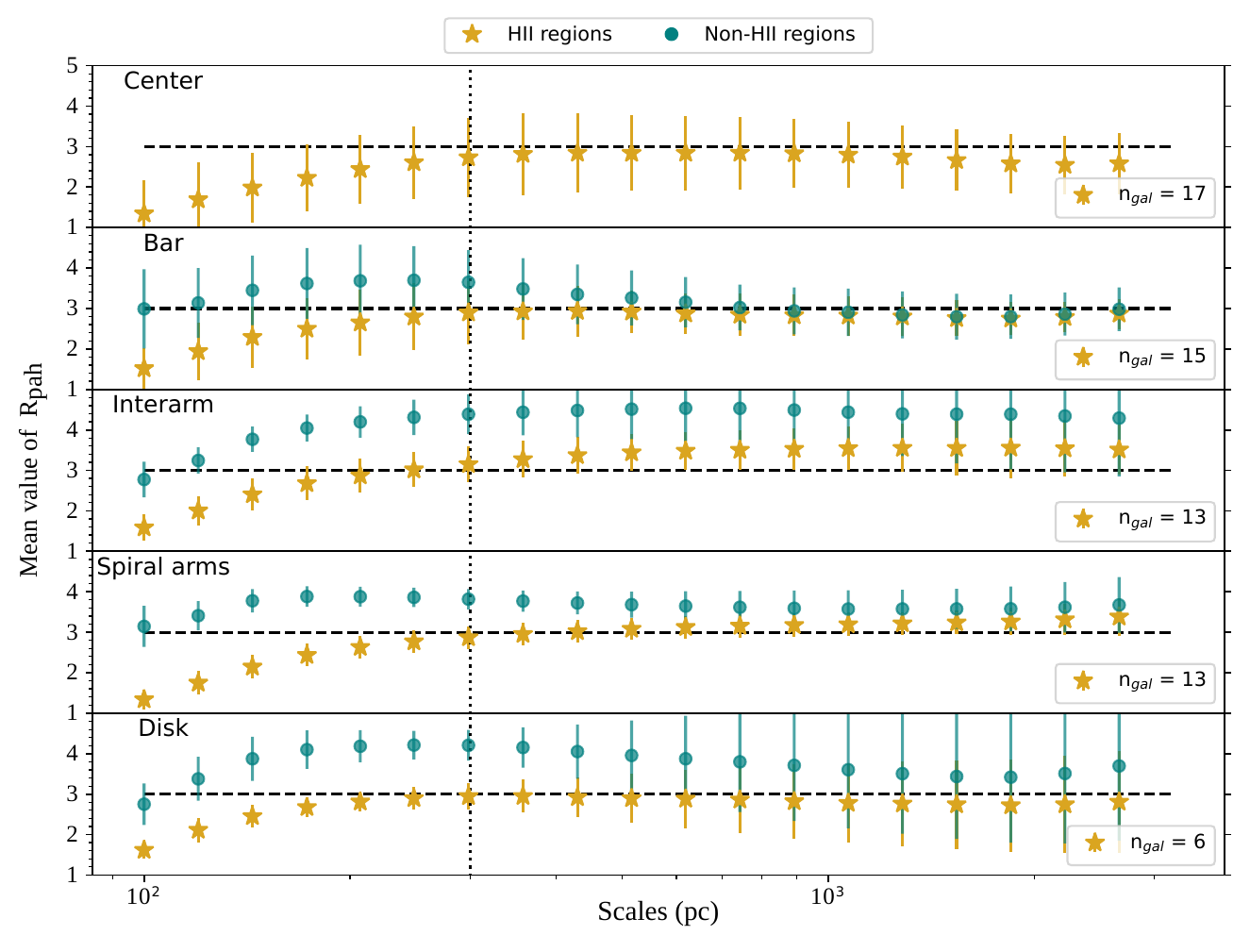}
    \caption{Mean of flux-weighted  \Rpah{} values at a given scale for each galaxy in different environments. \HII{} regions are indicated by a star and diffuse regions by a circle. Each point presents the mean \Rpah{} value for all the galaxies having this environment (written as n$_{\mathrm{gal}}$ in each plots) and the error bar presents the standard deviation of the \Rpah{} value at each scales.}
    \label{fig:rpah_flux_morph}
\end{figure*}

We look at flux weighted values of \Rpah{} in environmental regions (Figure~\ref{fig:rpah_flux_morph}). We find the same global trend of \Rpah{}, mostly for \HII{} regions. The transition to more stable values of \Rpah{} happens around $300$~pc, which is coherent with the identified transition scale in our analysis. We observe higher scatter at small scales between \HII{} and non-\HII{} regions than in the pixel-weighted average. This may once again be due to the higher flux in F2100W at small scales and overall in \HII{} regions (see Section~\ref{fluxwei_discussion}). Furthermore, for the interarm and disk regions, we find different values for \HII{} and non-\HII{} regions at large scales. While it is difficult to define \HII{} regions at these scales, in this work we only apply a mask for those regions, this might indicate that the F2100W emission is still high or contaminated in those regions at large scales.

\subsection{\texorpdfstring{\Rpah{}}{} stability in morphological regions}\label{app:rpah_stab}
\begin{figure*}[!htbp]
    \centering
    \includegraphics[width=13cm]{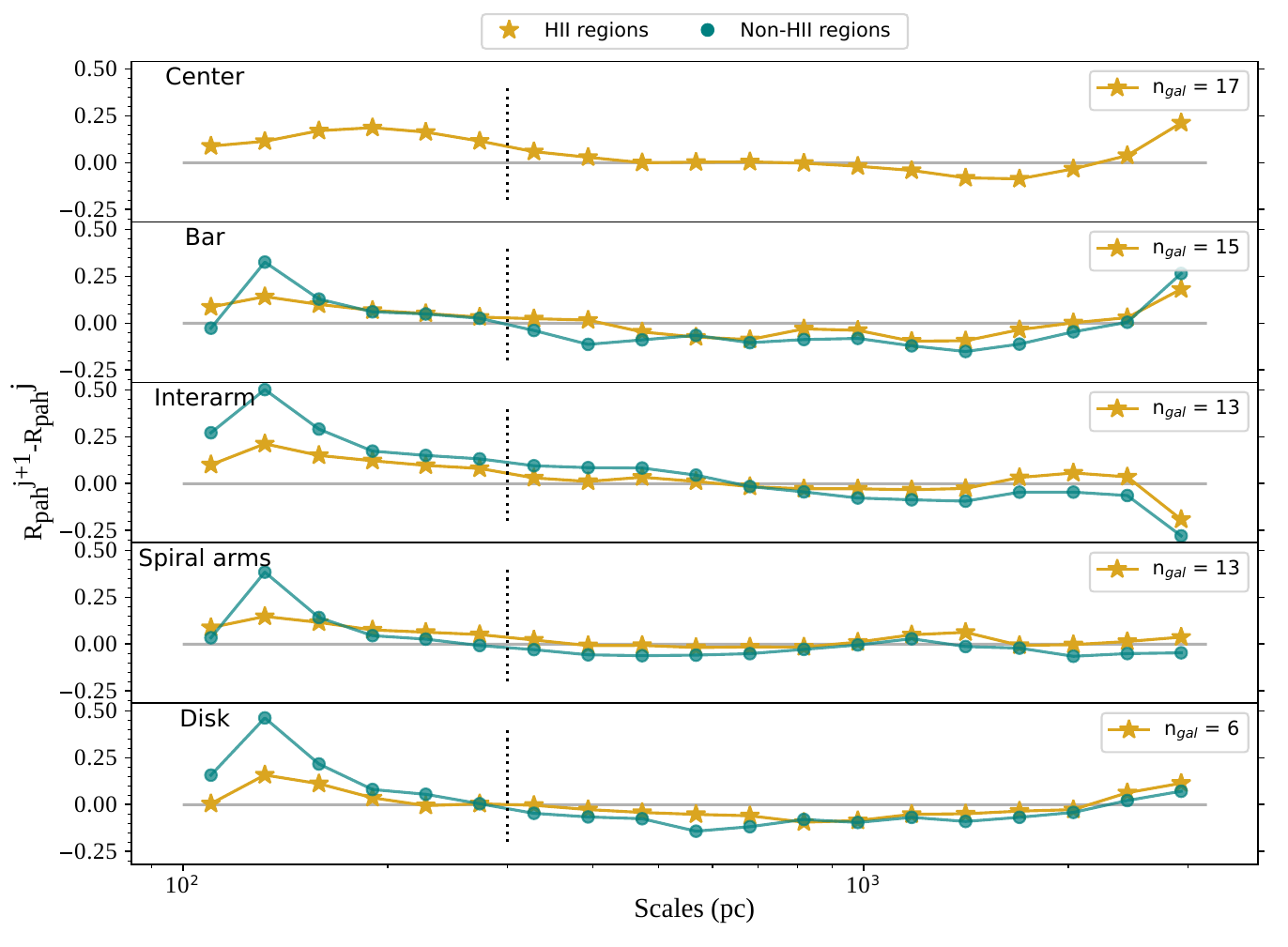}
    \caption{\Rpah{}$^{j+1}$-\Rpah{}$^{j}$ value at each scales for nebular and diffuse regions in different environments. \HII{} regions are indicated by a star and diffuse regions by a circle. The dashed line shows \Rpah{}$^{j+1}$-\Rpah{}$^{j} = 0$, corresponding to stable value of \Rpah{}. The dotted line show the transition scale of $300$~pc. }
    \label{fig:rpah_stab}
\end{figure*}

\begin{figure*}[!htbp]
    \centering
    \includegraphics[width=13cm]{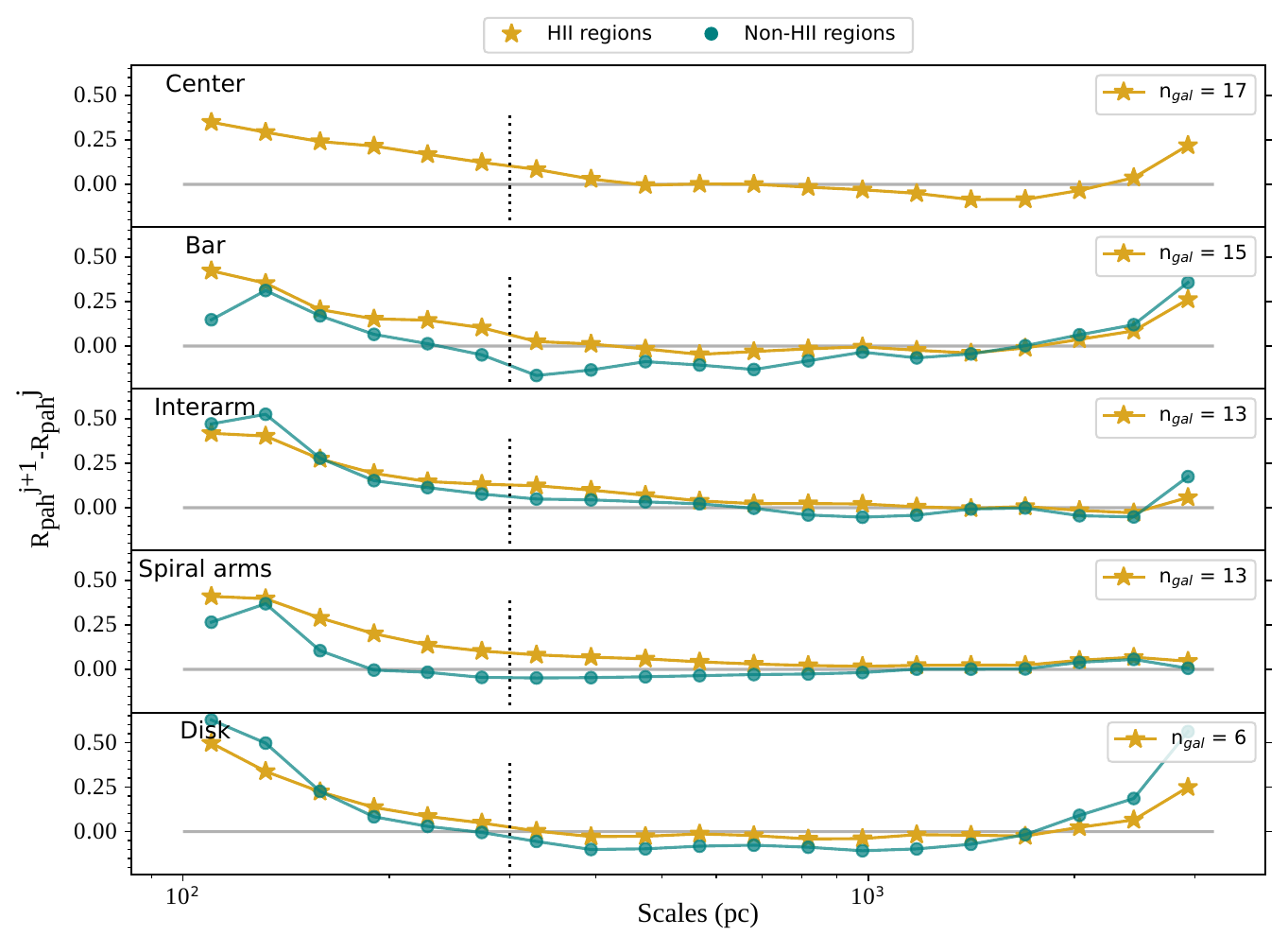}
    \caption{Flux-weighted \Rpah{}$^{j+1}$-\Rpah{}$^{j}$ value at each scales for nebular and diffuse regions in different environments. \HII{} regions are indicated by a star and diffuse regions by a circle. The dashed line shows \Rpah{}$^{j+1}$-\Rpah{}$^{j} = 0$, corresponding to stable value of \Rpah{}. The dotted line show the transition scale of $300$~pc. }
    \label{fig:rpah_stab_flux_wei}
\end{figure*}

We present in Figure \ref{fig:rpah_stab} and in Figure \ref{fig:rpah_stab_flux_wei} (for flux-weighted average) the two by two difference of \Rpah{} value at each scales, giving a way to measure the ``stability'' of \Rpah{} value. We observe that the values in \HII{} regions tends to be more stable than the value in diffuse regions. The transition scale also corresponds to the scale for which the \Rpah{} value stabilises. We observe that flux weighted values in \HII{} regions take longer to reach stability than in non-\HII{} regions.  

\end{appendix}
\end{document}